\documentclass[aps,superscriptaddress,preprintnumbers, superscriptaddress, showpacs, nofootinbibt,twocolumn, referee]{revtex4-1}
\usepackage{amsmath}
\usepackage{eurosym}
\usepackage{amssymb}
\usepackage{graphicx}
\usepackage{color}

\def\be{\begin{equation}}
\def\ee{\end{equation}}
\def\bea{\begin{eqnarray}}
\def\eea{\end{eqnarray}}

\begin{document}

\title{Series solution of the time-dependent Schr\"{o}dinger-Newton equations
in the presence of dark energy via the Adomian Decomposition Method}

\author{Tiberiu Harko}
\email{tiberiu.harko@aira.astro.ro}
\affiliation{Astronomical Observatory, 19 Ciresilor Street, 400487 Cluj-Napoca, Romania,}
\affiliation{Department of Physics, Babe\c s-Bolyai University, Mihail Kog\u alniceanu Street 1, 400084 Cluj-Napoca, Romania}
\affiliation{School of Physics, Sun Yat-Sen University, \\ Guangzhou 510275, People's
Republic of China,}
\author{Man Kwong Mak}
\email{mankwongmak@gmail.com}
\affiliation{Departamento de F\'{\i}sica, Facultad de Ciencias Naturales, Universidad de
Atacama, Copayapu 485, Copiap\'o, Chile.}
\author{Matthew J. Lake}
\email{matthewjlake@narit.or.th}
\affiliation{National Astronomical Research Institute of Thailand, \\ 260 Moo 4, T. Donkaew, A. Maerim, Chiang Mai 50180, Thailand}
\affiliation{Department of Physics and Materials Science, Faculty of Science, Chiang Mai University, \\ 239 Huaykaew Road, T. Suthep, A. Muang, Chiang Mai 50200, Thailand}
\affiliation{School of Physics, Sun Yat-Sen University, \\ Guangzhou 510275, People's Republic of China,}
\affiliation{Department of Physics, Babe\c s-Bolyai University, Mihail Kog\u alniceanu Street 1, 400084 Cluj-Napoca, Romania}
\affiliation{Office of Research Administration, Chiang Mai University,  239 Huaykaew Rd, T. Suthep, A. Muang, Chiang Mai 50200, Thailand}


\begin{abstract}
The Schr\"{o}dinger-Newton model is a nonlinear system obtained by coupling the linear Schr\"{o}dinger equation of canonical quantum mechanics with the Poisson equation of Newtonian mechanics.
In this paper we investigate the effects of dark energy on the time-dependent Schr\"{o}dinger-Newton equations by including a new source term with energy density $\rho_{\Lambda} = \Lambda c^2/(8\pi G)$, where $\Lambda$ is the cosmological constant, in addition to the particle-mass source term $\rho_m = m|\psi|^2$.
The resulting Schr\"{o}dinger-Newton-$\Lambda$ (S-N-$\Lambda$) system cannot be solved exactly, in closed form, and one must resort to either numerical or semianalytical (i.e., series) solution methods.
We apply the Adomian Decomposition Method, a very powerful method for solving a large class of nonlinear ordinary and partial differential equations, to obtain accurate series solutions of the S-N-$\Lambda$ system, for the first time. The dark energy dominated regime is also investigated in detail. We then compare our results to existing numerical solutions and analytical estimates, and show that they are consistent with previous findings. Finally, we outline the advantages of using the Adomian Decomposition Method, which allows accurate solutions of the S-N-$\Lambda$ system to be obtained quickly, even with minimal computational resources.
\newline
\\
\textbf{Keywords:} time-dependent Schr\"{o}dinger-Newton equations; dark
energy; Adomian Decomposition Method; series solutions
\end{abstract}

\pacs{04.50.Kd, 04.20.Cv, 04.20.Fy}
\maketitle
\tableofcontents

\section{Introduction}

Since the first attempts to create a quantum theory of the gravitational field
in the 1930's \cite{B1,B2}, the search for a complete theory of quantum gravity has been one of the
major fields of research in theoretical physics. In the pioneering work \cite%
{B1}, Bronstein investigated the quantum mechanical measurement of the $\Gamma
_{01}^0$ component of the Christoffel symbols.
This led to a fundamental limit on the temporal uncertainty intrinsic to any quantum measurement, $\Delta t\geq \left(\hbar /c^2G\rho ^2V\right)^{1/3}$, where $V$
and $\rho$ are the volume and the density of a self-gravitating
massive body, respectively. The time uncertainty can then be related to the spatial uncertainty via $\Delta
x \leq c\Delta t$. By introducing the standard mass-density-volume relation, $%
M = \rho V$, one also obtains the
mass-time-density uncertainty relation, $M \geq \hbar/c^2G\rho \left(\Delta
t\right)^3$.

This was one of the first generalised uncertainty relations (GURs) and, since then, many others have been proposed in the literature.
(See \cite{Tawfik:2015rva,Hossenfelder:2012jw} for reviews.)
Of these, the most widely studied are the generalised uncertainty principle (GUP) and the extended uncertainty principle (EUP).
The former aims to incorporate the effects of canonical gravitational attraction between quantum mechanical particles \cite{Adler:1999bu,Scardigli:1999jh}, while the latter accounts for the repulsive effects of dark energy in the form of a cosmological constant, $\Lambda$ \cite{Bolen:2004sq,Park:2007az,Bambi:2007ty}.
The extended generalised uncertainty principle (EGUP) accounts for both \cite{Lake:2018zeg,Lake:2019nmn,Lake2020-2}.

Over the years, many theoretical models and diverse approaches to the problem of quantum gravity have been developed. 
These include postulating the existence of the
graviton, the hypothetical spin-2 boson that mediates quantum gravitational interactions, string theory,
loop quantum gravity, and noncommutative geometry, to mention just a few of
the directions investigated.
(For detailed presentations of the different approaches, see \cite{Book1,Book2,Book3,Becker}.
For recent reviews of the present
status of quantum gravity research, see \cite{rev1,rev2,rev3,rev4}.)
However, recently, important progress in experimental techniques has enabled researchers to cool, control, and
measure physical systems in the weak gravity and quantum mechanical regimes, with far greater accuracy than ever before.
For the first time, it may be possible to directly observe quantum gravity effects at scales
accessible in terrestrial and near-Earth-orbit laboratories, in near-future experiments \cite{Carney,Aspelmeyer-1}.

Nonetheless, the conceptual and technical challenges to the construction of complete theory are manifold.
From a quantum field theory perspective, even the Newtonian theory of gravity is problematic.
Comparing Newton's law of gravity, $\Phi(r)=-Gm/r$, with Coulomb's law of electrostatics, $V(r)=k_e q/r$, it follows that the gravitational constant has mass dimension $-2$.
Hence, the theory of Newtonian quantum gravity is nonrenormalizable.
This result follows from the calculation of the graviton-graviton scattering at energy $E$, in which the divergent series $\sim \left[1+GE^2+\left(GE^2\right)^2+...\right]$ appears \cite{Zee}.
The nonrenormalizability of such naive quantum gravity theories therefore suggests that new physics should emerge at the Planck scale, $M_{\rm Pl}=\sqrt{\hbar c/G} \simeq 10^{19}m_{\rm proton}$.

Because of these challenges, much research interest has been devoted to the study of semi-classical models.
In these models matter fields are quantised while gravity remains classical, or is perturbatively quantised at next-to-leading order in the expansion of the metric.
Such an approach to quantum gravity was proposed in \cite{Fol}, which is based on the decomposition of the quantum metric into a classical and a fluctuating part, $\hat{g}_{\mu\nu}=g_{\mu\nu}+\delta\hat{g}_{\mu\nu}$.
Further assuming that $\langle\delta \hat{g}_{\mu\nu}\rangle=K_{\mu\nu}\ne 0$, where $K_{\mu\nu}$ is a classical tensor, one arrives at an effective gravitational Lagrangian of the form
$\mathcal{L}=\mathcal{L}_g\left(\hat{g}_{\mu\nu}\right)+
\sqrt{-g}\mathcal{L}_m\left(\hat{g}_{\mu\nu}\right)\simeq \mathcal{L}_g+\frac{\delta\mathcal{L}_g}{\delta g^{\mu\nu}}\delta\hat{g}^{\mu\nu}+
\sqrt{-g}\mathcal{L}_m+
\frac{\delta(\sqrt{-g}\mathcal{L}_m)}{\delta g^{\mu\nu}}\delta\hat{g}^{\mu\nu}$,
where $\kappa ^2=8\pi G/c^4$.
The gravitational field equations obtained from this Lagrangian lead to theories that require geometry-matter coupling at the classical level.
The coupling is of the kind that also appears in the $f(R,T)$ type modified gravity theories \cite{Od, book} and the cosmological implications of effective field theories with fluctuating metric components were investigated in \cite{Liu}.

However, because gravity is in many ways different from the other fundamental forces, and due to the intrinsic difficulties in its quantization, some researchers have suggested that the gravitational field may be essentially classical, and that it should not and cannot be quantized \cite{Mol, Ros}.
But, even gravity is not quantum, ordinary matter is.
Hence, in order to describe the gravitational dynamics of quantum fields, one must still combine classical gravity with quantized matter.
In this scenario quantized matter is coupled to the classical gravitational field by replacing the classical energy momentum tensor, $\hat{T}_{\mu \nu}$, with the expectation value of the energy-momentum operator, $\left <\hat{T}_{\mu \nu}\right >$, in Einstein's field equations.
The expectation value is constructed by averaging with respect to an appropriately chosen quantum state, $\Psi$, yielding the semi-classical field equations \cite{Carl},
\be\label{gr1}
R_{\mu \nu}-\frac{1}{2}g_{\mu \nu}R=\frac{8\pi G}{c^4}\left<\Psi \right |\hat{T}_{\mu \nu}\left |\Psi \right>.
\ee

Equations ~(\ref{gr1}) can be also obtained from the variational principle $\delta S=\delta \left(S_g+S_{\psi}\right)=0$ \cite{Kibble}, where $S_g=\left(1/16\pi G\right)\int{R\sqrt{-g}d^4x}$ is the standard Hilbert-Einstein action of general relativity, while the quantum part of the action, $S_{\psi}$, is introduced in the form
\be\label{9snl}
S_{\Psi}=\int{\left[{\rm Im}\left \langle \dot{\Psi}|\Psi\right \rangle-\left \langle \Psi |\hat{H}|\Psi \right \rangle +\alpha \left(\left \langle \Psi |\Psi \right \rangle -1\right) \right]dt}.
\ee
By varying the quantum action (\ref{9snl}) with respect to $\Psi$ we obtain the normalization condition $\left<\Psi|\Psi\right>=1$ and the following Schr\"odinger equation for $\Psi $,
\be\label{Sch}
i\hbar\left|\dot{\Psi}(t)\right>=\hat{H}(t)\left|\Psi (t)\right>-\alpha (t)\left|\Psi (t)\right>,
\ee
in addition to the semi-classical Einstein equations.
Note that the Bianchi identities still impose the conservation of the effective energy-momentum tensor, $\nabla _{\mu}\left<\Psi \right|\hat{T}^{\mu \nu}\left|\Psi \right>=0$.

The difficulty of building a successful quantum theory of general relativity, as well as the intrinsic problems of treating quantum field theories
in curved spacetimes, have also led to the hypothesis that a satisfactory description of quantum gravity could be achieved by unifying quantum mechanics with Newtonian gravity \cite{Diosi1}.
This corresponds to the weak field limit of Eqs. (\ref{gr1}), which reduce to the semi-classical Poisson equation
\begin{equation}\label{SN1}
\nabla^2 \Phi \left( \vec{r}\right) =4\pi G\left<\hat{\rho} \left(\vec{r}\right)\right> .
\end{equation}
Equation (\ref{SN1}) is the basis of the Schr{\" o}dinger-Newton approach \cite{Carney} and, in this model, the equation of motion of a self-gravitating massive particle can be formulated as
\begin{equation}\label{SG}
i\hbar \frac{\partial }{\partial t}\left|\psi \right>=\left(-\frac{\hbar^2}{2m}\nabla^2 + V+\Phi\right)\left|\psi \right> ,
\end{equation}
where $V$ is the canonical quantum potential and $\Phi$ is the gravitational self-interaction potential, obtained by solving (\ref{SN1}).

For a system of $N$ non-relativisitc free particles $V=0$ and the mass-density operator may be written as $\hat{\rho} = \sum_{i=1}^{N}m_i |\vec{r}-\vec{r}_{i}\rangle\langle\vec{r}-\vec{r}_{i}|$, where $\langle\vec{r}{\, '}-\vec{r}_{i}|\vec{r}-\vec{r}_{i}\rangle = \delta(\vec{r}{\, '}-\vec{r})$.
Equations (\ref{SN1})-(\ref{SG}) then combine to form
\begin{eqnarray}
i\hbar \frac{\partial \psi \left( \vec{r},t\right) }{\partial t} &=&-\frac{%
\hbar ^{2}}{2m}\nabla ^{2}\psi \left( \vec{r},t\right) \notag \\
&-&Gm^{2}\int \frac{\left\vert \psi \left( \vec{r}\;^{\prime },t\right)\right\vert ^{2}}{\left\vert \vec{r}-\vec{r}\;^{\prime }\right\vert} d{\vec{r}\;^{\prime }} \psi\left( \vec{r},t\right).
\end{eqnarray}
This is the standard Schr{\"o}dinger-Newton equation, whose static and time-dependent solutions have been intensively investigated in \cite{SN1,SN2,SN3,SN4,SN5,SN6,SN7,SN8,SN9,SN90, SN9a,SN10,SN11,SN12,SN13, SN14,SN15,SN16,SN17,SN18,SN19}.
The average value of the self-interaction potential can be obtained as $%
\left\langle \Phi \right\rangle \simeq \sqrt{\hbar ^{2}G/mR^{3}}$ and it
turns out that the particle behavior is essentially quantum if the condition $%
m^{3}R \ll \hbar ^{2}G\simeq 10^{-47}\;\mathrm{cm\;g^{3}}$, where $R$ is average radius of the wave function, is satisfied  \cite{Diosi1}.

Due to its extreme nonlinearity, the time-dependent Schr\"odinger-Newton system has mostly been investigated numerically.
An exception is the variational approach, which was considered in \cite{SN9a}, where the system of equations (\ref{SN1})-(\ref{SG}) was investigated in the hydrodynamical representation of quantum mechanics.
In this formalism the wave function is represented as $\psi \left(\vec{r},t\right)=\sqrt{\rho} \, e^{iS}$ and the canonical Schr\"odinger equation reduces to the equations of classical fluid mechanics, in the presence of the quantum potential.
The quantum fluid flows with a velocity $\vec{u}=\nabla S$ and the equations of motion can be obtained from the Lagrangian
\be
L\left(\rho,S,V\right)=\frac{\rho}{2}\left(\nabla S\right)^2+\rho \frac{\partial S}{\partial t}+\frac{\left(\nabla \rho\right)^2}{8\rho}+\frac{\left(\nabla V\right)^2}{8\pi}+\rho V.
\ee

By adopting a spherical Gaussian profile for the density, $\rho (r,t)=\pi^{-3/2}R^{-3}(t)e^{-r^2/R^2(t)}$, one can then obtain the gravitational potential and the Lagrangian of the system reduces to $L(R,\dot{R})=\dot{R}^2/2-1/2R^2+C/R$, where C is an arbitrary constant. The corresponding equation of motion for $R$ is $\ddot{R}=1/R^3-C/R^2$.
Using this formalism, one can obtain the energy eigenvalues, linear frequencies, and nonlinear late-time behavior of the S-N wave packet \cite{SN9a}.

More recently, the Schr\"{o}dinger-Newton system was generalized by considering the effects of dark energy in the form of a cosmological constant $\Lambda$ \cite{Matt}.
This is consistent with the standard $\Lambda$CDM model of cosmology, in which it is assumed that the late-time acceleration of the Universe is driven by a constant vacuum energy density,
$\rho_{\Lambda} = \Lambda c^2/(8\pi G) \simeq 10^{-30} \, {\rm g \;  cm^{-3}}$ \cite{Am}.
(For alternative models of dark energy as modified gravity, see \cite{RHL} and references therein.)
The physically interesting regime in which dark energy dominates both gravitational self-attraction and canonical quantum diffusion was investigated numerically and using analytical estimates.
It turns out that this takes place for objects with arbitrary mass that are sufficiently delocalized.
An estimate of the minimum delocalization width required, of the order of 67 m, was determined, and this prediction was verified by the numerical results.

However, the exact delocalisation radius required for dark energy domination can be much higher for very massive particles.
In general, the wave function of a free particle in the S-N-$\Lambda$ system was found to split into a core region that collapses due to gravitational self-attraction and an outer region that undergoes accelerated diffusion due to presence of dark energy \cite{Matt}.
While the former behaviour is present in the standard S-N model, the latter is unique to the S-N-$\Lambda$ system.
The order of magnitude of the critical radius separating collapse from expansion was found to match analytical estimates of the classical turnaround radius for a massive compact object in the presence of a cosmological constant \cite{r_TU}.

The goal of this paper is to further investigate the mathematical and
physical properties of the time-dependent S-N-$\Lambda$ system introduced in \cite{Matt}.
In order to obtain a better understanding of the dynamics of the model, we adopt a semi-analytical approach and construct series solutions using the Adomian Decomposition Method (ADM) \cite{new2, R1, R2,b2}.
This is a powerful method that can be used to obtain accurate series solutions of a large class of nonlinear differential equations, and systems of equations, with applications in diverse fields of science and engineering \cite{b5,b6,a0,a1,a11, a2,a20, a3,a4,a5,a6,a7,a8}.
Here, we apply it to the S-N-$\Lambda$ model for the first time.

An essential advantage of this method is that it can be used to obtain analytical approximations to the full numerical solutions, without any need for perturbation theory, closure approximations, linearization, or discretization methods.
For many highly nonlinear models, including the S-N and S-N-$\Lambda$ systems of equations, the use of these methods leads to complicated and time-consuming numerical computations.
On the other hand, to obtain even approximate closed-form analytical solutions of a nonlinear problem requires introducing restrictive and simplifying assumptions.
The key advantage of the ADM is that it can be used to find the solution of a given equation or system of equations in the form of a rapidly converging power series.
Successive terms in the series are obtained via a recursive relation, with the help of a special class of functions known as Adomian polynomials \cite{new2, R1, R2,b2}.
In most cases the series converges {\it fast}, so that the application of this method saves a lot of computational time.

Although the ADM has been used extensively in many areas of engineering and physics, it has been used very little in the study of gravitation and quantum mechanics.
(For some applications of the method in these fields, see \cite{a0,a20,a4}.)
In order to apply the method, we must first reformulate the time-dependent S-N-$\Lambda$ system as a system of
two integral equations.
We then obtain the series solutions of the system by expanding the nonlinear
terms using the Adomian polynomials \cite{new2,R1,R2,b2}.
To eliminate the unwanted oscillatory behavior of the solution, we represent the Adomian series in terms of their Pad\'{e} approximants.

After obtaining the recursive relation for the full S-N-$\Lambda$ system, we test the efficiency of the ADM for a free Gaussian wave packet, in the limit $G \rightarrow 0$, $\Lambda \rightarrow 0$.
In this case, the canonical Schr\"{o}dinger equation can be solved exactly, and we show that the ADM recovers the exact solution in just a few simple steps.
Next, series solutions are obtained for both the wave function and the gravitational potential, in the presence of gravitational self-interaction and dark energy.
The associated probability density is computed with the help of the Pad\'{e} approximants and we pay special attention to he dark energy dominated regime.

This paper is organized as follows. In Section~\ref{sect2} we
present the basic structure and mathematical formalism of the Adomian Decomposition Method.
The S-N-$\Lambda$ system is reformulated as a system of integral equations in Section~\ref{sect3}, and the recurrence relations for the series solution of the system are obtained.
The method is tested for the case of the canonical Schr\"{o}dinger equation describing the free propagation of a Gaussian wave packet, and it is shown that the exact solution of this system can be re-obtained in a few simple steps.
We obtain the semi-analytical solution of the time-dependent S-N-$\Lambda$ system, for Gaussian initial conditions, in Section~\ref{sect4}.
The dark energy dominated regime is also considered in detail, and a numerical analysis of the evolution of the probability density is presented. 
Our results are compared with previous analytical and numerical studies in Sec. \ref{sect5} and our discussion and final conclusions are presented in Section~\ref%
{sect6}.

\section{The Adomian Decomposition Method}\label{sect2}

Let us consider a partial differential equation written in the general form
\begin{equation}
\hat{L}_{t}\left[ u\left( x,t\right) \right] +\hat{R}\left[ u\left(
x,t\right) \right] +\hat{N}\left[ u\left( x,t\right) \right] =g(x,t),
\label{0}
\end{equation}%
where $\hat{L}_{t}\left[ .\right] =\partial /\partial t\left[ .\right] $, $%
\hat{R}\left[ .\right] $ is the linear remainder operator that may contain
partial derivatives with respect to $x$, $\hat{N}\left[ .\right] $ is a
nonlinear operator, which we assume is analytic, and $g$ is a non-homogeneous
term that is independent of $u$. Equation (\ref{0}) must be solved with the
initial condition $u(x,0)=f(x)$. We assume that $\hat{L}_{t}$ is invertible,
so that we can apply $\hat{L}_{t}^{-1}$ to both sides, obtaining
\begin{eqnarray}
u(x,t)&=&f(x)+\hat{L}_{t}^{-1}\left[ g(x,t)\right]
\notag \\
&-&\hat{L}_{t}^{-1}\hat{R}%
\left[ u\left( x,t\right) \right] - \hat{L}_{t}^{-1}\hat{N}\left[ u\left( x,t\right) \right] .
\end{eqnarray}

The ADM posits the existence of a series solution in which $u(x,t)$ is given by
\begin{equation}
u(x,t)=\sum_{n=0}^{\infty }u_{n}\left( x,t\right) ,  \label{01}
\end{equation}%
while the nonlinear term $\hat{N}\left[ u\left( x,t\right) \right] $ is
decomposed as
\begin{equation}
\hat{N}\left[ u\left( x,t\right) \right] =\sum_{n=0}^{\infty }A_{n}\left(
u_{0},u_{1},...,u_{n}\right) ,  \label{02}
\end{equation}%
where $\left\{ A_{n}\right\} _{n=0}^{\infty }$ are the Adomian polynomials.
These are generated according to the rule
\begin{equation}
A_{n}\left( u_{0},u_{1},...,u_{n}\right) =\frac{1}{n!}\frac{d^{n}}{d\epsilon
^{n}}\hat{N}\left(t,\sum_{k=0}^{n}\epsilon^{k}u_{k}\right)
\Bigg|_{\epsilon =0}.
\end{equation}

Substituting the series expansions (\ref{01}) and (\ref{02}) into Eq. (\ref%
{0}), we find
\begin{eqnarray}
\sum_{n=0}^{\infty }u_{n}\left( x,t\right) &=&f(x)+\hat{L}_{t}^{-1}\left[
g(x,t)\right] \notag \\
&-&\hat{L}_{t}^{-1}\hat{R}\left[ \sum_{n=0}^{\infty }u_{n}\left( x,t\right) %
\right] \notag \\
&-&\hat{L}_{t}^{-1}\left[ \sum_{n=0}^{\infty }A_{n}\left(
u_{0},u_{1},...,u_{n}\right) \right] .
\end{eqnarray}
Hence, we can obtain the following recurrence relation, giving the series
solution of Eq. (\ref{0}) as
\begin{equation}
u_{0}\left( x,t\right) =f(x)+\hat{L}_{t}^{-1}\left[ g(x,t)\right] ,
\end{equation}
\begin{eqnarray}
u_{k+1}(x,t)&=&\hat{L}_{t}^{-1}\hat{R}\left[ u_{k}\left( x,t\right) \right] -%
\hat{L}_{t}^{-1}\left[ A_{k}\left( u_{0},u_{1},...,u_{n}\right) \right],
\notag\\
&&k=0,1,2, \dots
\end{eqnarray}
Therefore, we obtain the approximate solution of Eq. (\ref{0}) as
\begin{equation}
u(x,t)\simeq \sum_{k=0}^{n}u_{k}\left( x,t\right) ,
\end{equation}
where
\begin{equation}
\lim_{n\rightarrow\infty }\sum_{k=0}^{n}u_{k}\left( x,t\right) =u(x,t).
\end{equation}

For a given nonlinearity $\hat{N}\left[ u%
\right] $, the Adomian polynomials are obtained as
\begin{equation}
A_{0}=\hat{N}\left[ u_{0}\right] , \quad A_{1}=u_{1}\frac{d}{du_{0}}\hat{N}\left[
u_{0}\right] ,
\end{equation}
\begin{equation}
A_{2}=u_{2}\frac{d}{du_{0}}\hat{N}\left[ u_{0}\right] +\frac{u_{1}^{2}}{2!}%
\frac{d^{2}}{du_{0}^{2}}\hat{N}\left[ u_{0}\right] ,
\end{equation}
\begin{equation}
A_{3}=u_{3}\frac{d}{du_{0}}\hat{N}\left[ u_{0}\right] +u_{1}u_{2}\frac{d^{2}%
}{du_{0}^{2}}\hat{N}\left[ u_{0}\right] +\frac{u_{1}^{3}}{3!}\frac{d^{3}}{%
du_{0}^{3}}\hat{N}\left[ u_{0}\right] ,
\end{equation}
and so on. The greater the number of terms, the higher the accuracy of the truncated series solution.

\section{The Adomian Decomposition Method for the time-dependent Schr\"{o}dinger-Newton-$\Lambda $ system}\label{sect3}

For a single particle of mass $m$, the time-dependent S-N-$\Lambda$ system is given by the following two equations,
\begin{equation}
i\hbar \frac{\partial \psi \left( \vec{r},t\right) }{\partial t}=-\frac{%
\hbar ^{2}}{2m}\nabla^2 \psi \left( \vec{r},t\right) +m\Phi \left( \vec{r}%
,t\right) \psi \left( \vec{r},t\right) ,  \label{1}
\end{equation}%
\begin{equation}
\nabla^2 \Phi \left( \vec{r},t\right) =4\pi Gm\left\vert \psi \left( \vec{r}%
,t\right) \right\vert ^{2}-\frac{1}{2}\Lambda c^{2},  \label{2}
\end{equation}%
where the last term in the above Poisson equation has been chosen so that the dark energy density is given by its standard form $\rho_{\Lambda} = \Lambda c^2/(8\pi G)$.
For spherically symmetric systems, Eqs. (\ref{1}) and (\ref{2}) take the form
\begin{equation}
i\hbar \frac{\partial \psi \left( r,t\right) }{\partial t}=-\frac{\hbar ^{2}%
}{2m}\frac{1}{r}\frac{\partial ^{2}}{\partial r^{2}}\left[ r\psi \left(
r,t\right) \right] +m\Phi \left( r,t\right) \psi \left( r,t\right) ,
\label{3}
\end{equation}%
\begin{equation}
\frac{\partial ^{2}}{\partial r^{2}}\left[ r\Phi \left( r,t\right) \right]
=4\pi Gmr\left\vert \psi \left( r,t\right) \right\vert ^{2}-\frac{1}{2}\Lambda c^{2}r,
\label{4}
\end{equation}
which must be solved with the boundary conditions $%
\psi \left( r,0\right) =\Psi (r)$ and $\Phi \left( r,t\right) =\phi \left(
r\right) $, respectively.

Introducing the operators $\hat{L}_{t}=\partial /\partial t$ and $\hat{L}%
_{rr}=\partial ^{2}/\partial r^{2}$, Eqs. (\ref{3}) and (\ref{4}) can be
rewritten as
\begin{equation}
\hat{L}_{t}\psi \left( r,t\right) =\frac{1}{i}\left\{ -\frac{\hbar }{2m}%
\frac{1}{r}\frac{\partial ^{2}}{\partial r^{2}}\left[ r\psi \left(
r,t\right) \right] +\frac{m}{\hbar }\Phi \left( r,t\right) \psi \left(
r,t\right) \right\} ,  \label{5}
\end{equation}
\begin{equation}
\hat{L}_{rr}\left[ r\Phi \left( r,t\right) \right] =4\pi Gmr\left\vert \psi
\left( r,t\right) \right\vert ^{2}-\frac{1}{2}\Lambda c^{2}r.  \label{6}
\end{equation}
These equations can be solved formally to give
\begin{eqnarray}
\hspace{-0.7cm}&&\psi \left( r,t\right) =\Psi (r)+  \notag \\
\hspace{-0.7cm}&&\frac{1}{i}\hat{L}_{t}^{-1}\left\{ -\frac{\hbar }{2m}\frac{1%
}{r}\frac{\partial ^{2}}{\partial r^{2}}\left[ r\psi \left( r,t\right) %
\right] +\frac{m}{\hbar }\Phi \left( r,t\right) \psi \left( r,t\right)
\right\} ,  \label{7}
\end{eqnarray}
\begin{equation}
r\Phi \left( r,t\right) =\hat{L}_{rr}^{-1}\left[ 4\pi Gmr\left\vert \psi
\left( r,t\right) \right\vert ^{2}-\frac{1}{2}\Lambda c^{2}r\right] .  \label{8}
\end{equation}

By taking into account the fact that $\hat{L}_{t}^{-1}\left( .\right)
=\int_{0}^{t}\left( .\right) dt$ and $\hat{L}_{rr}^{-1}=\int_{0}^{r}%
\int_{0}^{r}\left( .\right) drdr$, we then obtain
\begin{eqnarray}
\hspace{-0.9cm}&&\psi \left( r,t\right) =\Psi (r)+  \notag \\
\hspace{-0.9cm}&& \frac{1}{i}\int_{0}^{t}\left\{ -\frac{\hbar }{2m}\frac{1}{r%
}\frac{\partial ^{2}}{\partial r^{2}}\left[ r\psi \left( r,t\right) \right] +%
\frac{m}{\hbar }\Phi \left( r,t\right) \psi \left( r,t\right) \right\} dt,
\label{9}
\end{eqnarray}
\begin{eqnarray}  \label{10}
r\Phi \left( r,t\right)& =&\phi _{1}(t)r+\phi _{2}(t)  \notag \\
&+&4\pi Gm\int_{0}^{r}\int_{0}^{\xi }\xi ^{\prime }\left\vert \psi \left( \xi
^{\prime },t\right) \right\vert ^{2}d\xi ^{\prime }d\xi -\frac{\Lambda c^{2}%
}{12}r^{3},  \notag \\
\end{eqnarray}
where $\phi _{1}(t)$ and $\phi _{2}(t)$ are arbitrary integration functions.
Using the Cauchy formula for repeated integration, Eq.~(\ref{10}) can be rewritten as
\begin{eqnarray}  \label{11}
\Phi \left( r,t\right) &=&\phi _{1}(t)+\frac{\phi _{2}(t)}{r}  \notag \\
&+&4\pi Gm\int_{0}^{r}\xi \left( 1-\frac{\xi }{r}\right) \left\vert \psi
\left( \xi ,t\right) \right\vert ^{2}d\xi -\frac{\Lambda c^{2}}{12}r^{2}.
\notag \\
\end{eqnarray}

In order to make the gravitational potential finite in the origin $r=0$, we
chose $\phi _{2}(t)=0$, giving
\begin{equation}
\Phi \left( r,t\right) =\phi _{1}(t)+4\pi Gm\int_{0}^{r}\xi \left( 1-\frac{%
\xi }{r}\right) \left\vert \psi \left( \xi ,t\right) \right\vert ^{2}d\xi -%
\frac{\Lambda c^{2}}{12}r^{2}.  \label{12}
\end{equation}
We now determine the series solution of the reformulated system of equations, (\ref{9}) and (\ref{12}), by assuming expansions of the form
\begin{equation}
\psi (r,t)=\sum_{n=0}^{\infty }\psi {_{n}(r,t)},  \label{13}
\end{equation}%
and
\begin{equation}
\Phi \left( r,t\right) =\sum_{n=0}^{\infty }\Phi {_{n}(r,t).}  \label{14}
\end{equation}%
In addition, we decompose the terms $\Phi \left( r,t\right) \psi
\left( r,t\right) $ and $\left\vert \psi \left( \xi ,t\right) \right\vert
^{2}$ in terms of the Adomian polynomials as
\begin{eqnarray}  \label{15}
\hspace{-0.4cm}\Phi \left( r,t\right) \psi \left( r,t\right)
&=&\sum_{n=0}^{\infty }{A_{n}(r,t)},  \notag \\
\hspace{-0.4cm}\left\vert \psi \left( \xi ,t\right) \right\vert ^{2}&=&\psi
\left( \xi ,t\right) \psi ^{\ast }\left( \xi ,t\right) =\sum_{n=0}^{\infty }{%
B_{n}(\xi ,t)}.
\end{eqnarray}
Substituting the above decompositions into Eqs.~(\ref{9}) and (\ref{12}) we
obtain
\begin{widetext}
\begin{equation}
\sum_{n=0}^{\infty }\psi {_{n}(r,t)}=\Psi (r)+\frac{1}{i}\sum_{n=0}^{\infty
}\int_{0}^{t}\left\{ -\frac{\hbar }{2m}\frac{1}{r}\frac{\partial ^{2}}{%
\partial r^{2}}\left[ r\psi {_{n}(r,t)}\right] +\frac{m}{\hbar }{A_{n}(r,t)}%
\right\} dt,  \label{16}
\end{equation}%
\begin{equation}
\sum_{n=0}^{\infty }\Phi {_{n}(r,t)}=\phi _{1}(t)+4\pi Gm\sum_{n=0}^{\infty
}\int_{0}^{r}\xi \left( 1-\frac{\xi }{r}\right) {B_{n}(\xi ,t)}d\xi -\frac{%
\Lambda c^{2}}{12}r^{2}.
\end{equation}
\end{widetext}

Hence, we obtain the following recursive series solution for the
time-dependent S-N-$\Lambda $ system,
\begin{equation}
\psi _{0}(r,t)=\Psi (r),  \label{17}
\end{equation}%
\begin{eqnarray}  \label{18}
\psi _{k+1}(r,t)&=& \int_{0}^{t}\left\{ i\frac{\hbar }{2m}\frac{1}{r}\frac{%
\partial ^{2}}{\partial r^{2}}\left[ r\psi {_{k}(r,t)}\right] -\frac{m}{%
\hbar }{A_{k}(r,t)}\right\} dt,  \notag \\
&&k=0,1,2,...,n,
\end{eqnarray}%
\begin{equation}
\Phi {_{0}(r,t)}=\phi _{1}(t)-\frac{\Lambda c^{2}}{12}r^{2},  \label{19}
\end{equation}%
\begin{eqnarray}  \label{20}
\Phi _{k+1}(r,t)&=&4\pi Gm\int_{0}^{r}\xi \left( 1-\frac{\xi }{r}\right) {%
B_{k}(\xi ,t)}d\xi ,  \notag \\
&&k=0,1,2,..,n.
\end{eqnarray}
The first three Adomian polynomials in each series are obtained as
\begin{equation}
A_{0}{(r,0)}=\psi _0(r,0)\Phi {_{0}(r,0)},  \label{21}
\end{equation}
\begin{equation}
A_{1}{(r,t)}=\psi _{0}(r,0)\Phi {_{1}(r,t)}+\psi {_{1}(r,t)}\Phi {_{0}}(r,0),
\label{22}
\end{equation}
\begin{eqnarray}  \label{23}
A_{2}{(r,t)}&=&\psi _0(r,0)\Phi {_{2}(r,t)}+\psi _{1}(r,t)\Phi _{1}(r,t)+
\notag \\
&&\psi_{2}(r,t)\Phi _{0}(r,0),
\end{eqnarray}
and
\begin{equation}
B_{0}\left( r,0\right) =\psi _{0}\left( r,0\right) \psi _{0}^{\ast }\left(
r,0\right) ,  \label{24}
\end{equation}
\begin{equation}
B_{1}{(r,t)}=\psi _0(r,0)\psi _{1}^{\ast }(r,t)+\psi _{1}(r,t)\psi {%
_{0}^{\ast }(r,0)},  \label{25}
\end{equation}
\begin{eqnarray}  \label{26}
B_{2}{(r,t)}&=&\psi _0(r,0)\psi _{2}^{\ast }(r,t)+\psi _{1}(r,t)\psi
_{1}^{\ast }(r,t)+  \notag \\
&&\psi _{2}(r,t)\psi _{0}^{\ast }(r,0),
\end{eqnarray}
respectively.

For simplicity, we assume that the initial state of the
wave function is a spherically symmetric Gaussian,
\begin{equation} \label{initial_condition}
\psi (r,0)=\Psi \left( r\right) =\left( \frac{\alpha }{\pi }\right)
^{3/4}e^{-\alpha r^{2}}, \quad \alpha =\mathrm{constant}, 
\end{equation}%
with initial width $\sigma_0 =1/\sqrt{2\alpha }$.
The gravitational potential $\phi (r)$, corresponding to the
time-independent Gaussian wave packet, then satisfies the Poisson
equation
\begin{equation}
\frac{1}{r}\frac{d^{2}}{dr^{2}}\left[ r\phi (r)\right] =4\pi Gm\left( \frac{%
\alpha }{\pi }\right) ^{3/2}e^{-2\alpha r^{2}}-\frac{1}{2}\Lambda c^{2},
\end{equation}%
whose general solution is
\begin{equation}
\phi (r)=-\frac{1}{12}c^{2}\Lambda r^{2}-\frac{c_{1}}{r}+c_{2}-\frac{Gm\;%
\text{erf}\left( \sqrt{2\alpha }r\right) }{2\sqrt{2}r},
\end{equation}%
where $\text{erf}(z)=\left( 2/\sqrt{\pi }\right) \int_{0}^{z}{e^{-t^{2}}dt}$
is the error function, and $c_{1}$ and $c_{2}$ are arbitrary integration
constants. In order to avoid a singularity at the origin, we take $c_{1}=0$%
. The initial distribution of the gravitational potential is then finite
everywhere, and satisfies the condition $\lim_{r\rightarrow 0}\phi
(r)=c_{2}-Gm\sqrt{\alpha /\pi }$.

\subsection{Testing the Adomian Decomposition Method} \label{sect3.1}

To test the efficiency of the ADM, we
consider the evolution of a Gaussian wave packet in the absence of the
gravitational interaction, and dark energy, by setting $\Phi (r,t)=0$.
In this case, the evolution
of the wave function is given by the canonical Schr\"{o}dinger equation
\begin{equation}
\frac{\partial \psi \left( r,t\right) }{\partial t}=i\frac{\hbar }{2m}\frac{1%
}{r}\frac{\partial ^{2}}{\partial r^{2}}\left[ r\psi \left( r,t\right) %
\right] ,  \label{28}
\end{equation}%
and must be solved subject to the initial condition $\psi (r,0)=\left( \alpha
/\pi \right) ^{3/4}e^{-\alpha r^{2}}$ (\ref{initial_condition}). The general solution of Eq.~(\ref{28}%
), satisfying the required initial condition, is given by
\begin{equation}
\psi (r,t)=\left( \frac{\alpha }{\pi }\right) ^{3/4}\left[ \frac{1}{%
1+2i\left( \alpha \hbar /m\right) t}\right] ^{3/2}e^{-\frac{\alpha r^{2}}{%
1+2i\left( \alpha \hbar /m\right) t}}.
\end{equation}
The associated probability distribution $P\left( r,t\right) =\left\vert \psi (r,t)\right\vert ^{2}$  is obtained as
\begin{eqnarray}
P\left( r,t\right) =\left( \frac{%
\alpha }{\pi }\right) ^{3/2}\left[ \frac{1}{1+4\left( \alpha \hbar /m\right)
^{2}t^{2}}\right] ^{3/2}
e^{-\frac{\alpha r^{2}}{1+4\left( \alpha \hbar /m\right) ^{2}t^{2}}}.\nonumber\\
\end{eqnarray}

In order to simplify the mathematical formalism, we introduce a new set of
dimensionless variables $\left( \tau ,\theta \right) $, defined as
\begin{equation}  \label{51}
t=\frac{m}{\alpha \hbar }\tau =1.58\times 10^{-9}\times \left(\frac{m}{m_p}%
\right)\left(\frac{\alpha}{10^{12}\,\mathrm{cm ^{-2}}}\right)^{-1}\times
\tau \;\mathrm{s},
\end{equation}
where $m_p$ denotes the proton mass, and
\begin{equation}
r=\frac{1}{\sqrt{\alpha }}\theta=10^{-6}\times \left(\frac{\alpha}{10^{12}\;%
\mathrm{cm}^{-2}}\right)^{-1/2}\times \theta \;\mathrm{cm},
\end{equation}
respectively. Moreover, we rescale the wave function so that
\begin{equation}  \label{52}
\psi (r,t)=\left( \frac{\alpha }{\pi }\right) ^{3/4}\tilde{\psi}\left(
\theta ,\tau \right) .
\end{equation}

The rescaled wave function satisfies the dimensionless Schr\"{o}dinger
equation
\begin{equation}
\frac{\partial \tilde{\psi}\left( \theta ,\tau \right) }{\partial \tau }=%
\frac{i}{2}\frac{1}{\theta }\frac{\partial ^{2}}{\partial \theta ^{2}}\left[
\theta \tilde{\psi} \left( \theta ,\tau \right) \right] ,  \label{28a}
\end{equation}%
whose general solution, satisfying the initial condition (\ref{initial_condition}), is given
by
\begin{equation}
\tilde{\psi}\left( \theta ,\tau \right) =\left[ \frac{1}{1+2i\tau }\right]
^{3/2}e^{-\frac{\theta ^{2}}{1+2i\tau }}.
\end{equation}
This may be expanded as a power series, with respect
to the dimensionless time $\tau $, as
\begin{eqnarray}
\tilde{\psi}\left( \theta ,\tau \right) &\simeq &e^{-\theta ^{2}}+\left(
2\theta ^{2}-3\right) e^{-\theta ^{2}}\frac{i\tau }{1!}  \notag  \label{44}
\\
\hspace{-0.5cm} &+&\left[ 4\theta ^{2}\left( \theta ^{2}-5\right) +15\right]
e^{-\theta ^{2}}\frac{\left( i\tau \right) ^{2}}{2!}  \notag \\
&+&\left( 8\theta ^{6}-84\theta ^{4}+210\theta ^{2}-105\right)
e^{-\theta ^{2}}\frac{\left( i\tau \right) ^{3}}{3!}
\notag \\
&+& \dots
\end{eqnarray}

To solve Eq. (\ref{28a}) using the ADM, we apply the operator $\hat{L}_{\tau }^{-1}$ to both sides, giving
\begin{equation}
\tilde{\psi}\left( \theta ,\tau \right) =\tilde{\psi}\left( \theta ,0\right)
+\frac{i}{2}\hat{L}_{\tau }^{-1}\left\{ \frac{1}{\theta }\frac{\partial ^{2}%
}{\partial \theta ^{2}}\left[ \theta \tilde{\psi}\left( \theta ,\tau \right) %
\right] \right\} .  \label{29}
\end{equation}
We then decompose $\tilde{\psi}\left( \theta ,\tau \right) $ into an infinite
sum of components, so that $\tilde{\psi}\left( \theta ,\tau \right)
=\sum_{n=0}^{\infty }\tilde{\psi}_{n}\left( \theta ,\tau \right) $, where
the components $\tilde{\psi}_{n}\left( \theta ,\tau \right) $ will be
determined recurrently. By substituting the series expansion into Eq. (\ref%
{29}), we obtain
\begin{equation}
\sum_{n=0}^{\infty }\tilde{\psi}_{n}\left( \theta ,\tau \right) =\tilde{\psi}%
\left( \theta ,0\right) +\frac{i}{2}\hat{L}_{\tau }^{-1}\sum_{n=0}^{\infty
}\left\{ \frac{1}{\theta }\frac{\partial ^{2}}{\partial \theta ^{2}}\left[
\theta \tilde{\psi}_{n}\left( \theta ,\tau \right) \right] \right\} .
\end{equation}
Thus, we obtain the recursive relations
\begin{equation}
\tilde{\psi}_{0}\left( \theta ,0\right) =\tilde{\psi}\left( \theta ,0\right)
,
\end{equation}%
\begin{equation}
\tilde{\psi}_{n+1}(\theta ,\tau )=\frac{i}{2}\hat{L}_{\tau }^{-1}\left\{
\frac{1}{\theta }\frac{\partial ^{2}}{\partial \theta ^{2}}\left[ \theta
\tilde{\psi}_{n}\left( \theta ,\tau \right) \right] \right\} , \, n=0,1,2,...
\end{equation}
The first few iterations are given by
\begin{eqnarray}
\tilde{\psi}_{1}(\theta ,\tau )&=&\frac{i}{2}\hat{L}_{\tau }^{-1}\left\{
\frac{1}{\theta }\frac{\partial ^{2}}{\partial \theta ^{2}}\left[ \theta
\tilde{\psi}{_{0}(\theta ,0)}\right] \right\} \notag \\
&=&\left( 2\theta ^{2}-3\right) e^{-\theta ^{2}}\frac{i\tau }{1!},
\end{eqnarray}%
\begin{eqnarray}
\tilde{\psi}_{2}(\theta ,\tau )&=&\frac{i}{2}\hat{L}_{\tau
}^{-1}\left\{ \frac{1}{\theta }\frac{\partial ^{2}}{\partial \theta ^{2}}%
\left[ \theta \tilde{\psi}{_{1}(\theta ,\tau )}\right] \right\} \notag \\
&=&\left[ 4\theta ^{2}\left( \theta ^{2}-5\right) +15\right]
e^{-\theta ^{2}}\frac{\left( i\tau \right) ^{2}}{2!},
\end{eqnarray}
and
\begin{eqnarray}
\tilde{\psi}_{3}(\theta ,\tau )&=&\frac{i}{2}%
\hat{L}_{\tau }^{-1}\left\{ \frac{1}{\theta }\frac{\partial ^{2}}{\partial
\theta ^{2}}\left[ \theta \tilde{\psi}{_{2}(\theta ,\tau )}\right] \right\}
\notag \\
&=&\left( 8\theta ^{6}-84\theta ^{4}+210\theta ^{2}-105\right) e^{-\theta
^{2}}\frac{\left( i\tau \right) ^{3}}{3!},
\end{eqnarray}
and so on. Clearly, the Adomian series solution $\tilde{\psi}\left( \theta ,\tau \right) \simeq
\tilde{\psi}_{0}\left( \theta ,0\right) +\tilde{\psi}_{1}\left( \theta ,\tau
\right) +\tilde{\psi}_{2}\left( \theta ,\tau \right) +\tilde{\psi}_{3}\left(
\theta ,\tau \right) + \dots$ exactly reproduces the series expansion of the exact solution (\ref{44}) and, in the
limit of an infinite number of iterations, fully recovers it. Hence, we
have shown that the ADM gives the exact series
representation of the solution of the spherically symmetric, three-dimensional Schr\"{o}dinger equation describing the time-evolution of a free
Gaussian wave packet.

The probability distribution $P(\theta,\tau)$ is obtained as $%
P(\theta,\tau)=\lim_{n\rightarrow \infty}\left[\left(\sum _{k=0}^n{\psi
_k(\theta, \tau)}\right)\left(\sum _{k=0}^n{\psi _k^{*}(\theta, \tau)}\right)%
\right]=(\alpha /\pi)^{3/2}\lim_{n\rightarrow \infty}\left[\left(\sum
_{k=0}^n{\tilde{\psi} _k(\theta, \tau)}\right)\left(\sum _{k=0}^n{\tilde{\psi%
} _k^{*}(\theta, \tau)}\right)\right]=(\alpha /\pi)^{3/2}\tilde{P}%
(\theta,\tau)$.
Here, we approximate the series representation of $\tilde{P}%
(\theta,\tau)$ by its Pad\'{e} approximant $\tilde{P}[3/4](\theta, \tau)$,
which is given by
\begin{widetext}
\be\label{63a1}
\tilde{P}\left[3/4\right](\theta, \tau)=\frac{e^{-2 \theta ^2}+\frac{2 e^{-2 \theta ^2} \left(64 \theta ^6+48 \theta ^4-12 \theta
   ^2+3\right) \tau ^2}{3 \left(16 \theta ^4-8 \theta ^2+3\right)}}{-\frac{4 \left(64 \theta
   ^6-144 \theta ^4+60 \theta ^2-15\right) \tau ^2}{3 \left(16 \theta ^4-8 \theta
   ^2+3\right)}+\frac{2 \left(256 \theta ^8-768 \theta ^6+864 \theta ^4-240 \theta
   ^2+45\right) \tau ^4}{3 \left(16 \theta ^4-8 \theta ^2+3\right)}+1}.
\ee
\end{widetext}
Generally, for a power series of the form $f(z)=\sum_{z=0}^{\infty}f_kz^k$ the Pad\'{e} approximant of the order $(m,n)$ in the vicinity of the point $z=0$ is the rational function $\Pi_{m.n}\in R_{m,n}$ having the property that it takes the closest values to the given series near $z=0$. Here, by $R_{m,n}$, we have denoted the set of rational functions of the form $P/Q$, where $P$ and $Q$ are polynomials in $z$ of degree $p\leq m$ and $q\leq n$, respectively \cite{Pade}.

\begin{figure}[h!]
\centering
\includegraphics[width=7cm]{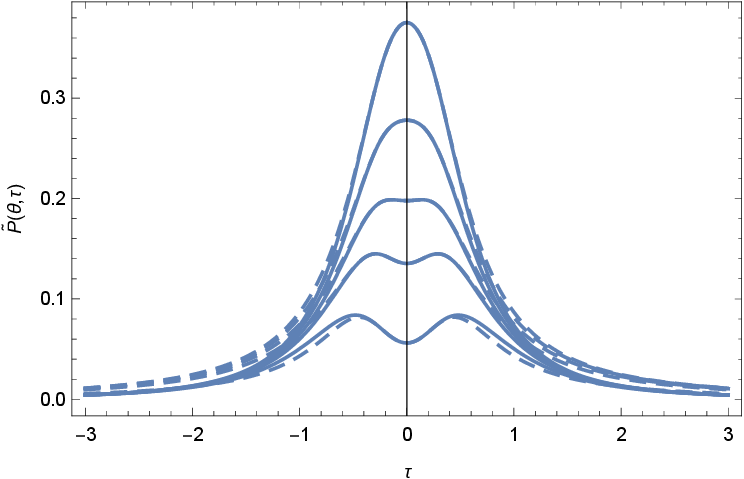}
\caption{Comparison of the Pad\'{e} approximants of the ADM
series solutions and exact solutions of the Schr%
\"{o}dinger equation describing the probability density of the free evolution of a Gaussian wave packet, $\tilde{P}(\protect\theta,
\protect\tau)$, for $%
\protect\theta =1.2$ (lower curve), $\protect\theta =1$, $\protect\theta %
=0.9 $, $\protect\theta =0.8$, and $\protect\theta =0.7$ (upper curve),
respectively. The exact solutions are represented by solid curves, while the
Pad\'{e} approximants of the Adomian series are plotted as dashed curves.}
\label{fig1}
\end{figure}

The comparison between the exact probability density of the Gaussian quantum
wave packet,
\be\label{thetap0}
\tilde{P}(\theta,\tau)=\frac{e^{-2\theta ^2/\left(1+4\tau ^2\right)}}{\left(1+4\tau ^2\right)^{3/2}},
\ee
and its approximation, given by Eq.~(\ref{63a1}), is represented
in Fig.~\ref{fig1}.
As one can see from this figure, Eq.~(\ref{63a1}) gives an excellent
description of the time-evolution of the wave packet for $-1\leq
\tau \leq 1$, and a good approximation for $\tau$
outside this range.

\section{Series solution of the time-dependent Schr\"{o}dinger-Newton-$%
\Lambda $ system}\label{sect4}

Using the mathematical formalism developed in the previous Section, we now
construct explicit series solutions of the S-N-$\Lambda $ system.
In addition, we perform a detailed numerical
study. 
Our main goal is to highlight the effects
of self-gravity and dark energy on the time-evolution of a the quantum wave
packet.

\subsection{Semi-analytical solutions} \label{sect3.2}

In terms of the dimensionless variables $(\tau ,\theta)$ and rescaled wave function $\tilde{\psi}(\tau ,\theta)$,
defined in Eqs.~(\ref{51}) and (\ref{52}), the time-dependent S-N-$\Lambda $ equations take the form
\begin{equation}  \label{63a}
\frac{\partial \tilde{\psi}(\theta ,\tau )}{\partial \tau }=i\left\{ \frac{1%
}{2\theta }\frac{\partial ^{2}}{\partial \theta ^{2}}\left[ \theta \tilde{%
\psi}(\theta ,\tau )\right] -\tilde{\Phi}(\theta ,\tau )\tilde{\psi}(\theta
,\tau )\right\} ,
\end{equation}%
\begin{equation}  \label{64a}
\frac{\partial ^{2}}{\partial \theta ^{2}}\left[ \theta \tilde{\Phi}(\theta
,\tau )\right] =\sigma \theta \left\vert \tilde{\psi}(\theta ,\tau
)\right\vert ^{2}-\lambda \theta ,
\end{equation}%
where we have defined the dimensionless parameters
\begin{equation}  \label{63}
\sigma =\frac{4Gm^{3}}{\sqrt{\alpha \pi }\hbar ^{2}}=6.31\times
10^{-31}\times \left(\frac{m}{m_p}\right)^3\left(\frac{\alpha }{10^{12}\;%
\mathrm{cm}^{-2}}\right)^{-1/2},
\end{equation}%
\begin{eqnarray}\label{63l}
\lambda =\frac{m^{2}c^{2}}{2\alpha ^{2}\hbar ^{2}}\Lambda &=&1.125\times
10^{-53}\times \left(\frac{m}{m_p}\right)^2 \notag \\
&\times &\left(\frac{\alpha}{10^{12}\;\mathrm{cm}^{-2}}\right)^{-2}\left(\frac{%
\Lambda}{10^{-56}\;\mathrm{cm}^{-2}}\right),
\end{eqnarray}
and introduced the rescaled gravitational potential as
\begin{eqnarray}  \label{64}
\tilde{\Phi}(\theta ,\tau )&=&\frac{m^{2}}{\alpha ^2\hbar ^{2}}\Phi (\theta
,\tau )=2.51\times 10^{-18} \notag \\
&\times &\left(\frac{m}{m_p}\right)^2\left(\frac{\alpha}{10^{12}\;\mathrm{cm}^{-2}}%
\right)^{-2} \Phi (\theta ,\tau ).
\end{eqnarray}

The solution of Eqs.~(\ref{63a}) and (\ref{64a}) can be represented as
\begin{eqnarray}
\tilde{\psi}(\theta ,\tau )&=&\tilde{\Psi}\left( \theta \right) \notag \\
&+&i\int_{0}^{\tau }\left\{ \frac{1}{2\theta }\frac{\partial ^{2}}{\partial
\theta ^{2}}\left[ \theta \tilde{\psi}(\theta ,\tau )\right] -\tilde{\Phi}%
(\theta ,\tau )\tilde{\psi}(\theta ,\tau )\right\} d\tau ,  \notag \\
\end{eqnarray}
\begin{equation}
\tilde{\Phi}(\theta ,\tau )=\tilde{\phi}_{1}(\tau )+\sigma \int_{0}^{\theta
}\xi \left( 1-\frac{\xi }{\theta }\right) \left\vert \tilde{\psi}\left( \xi
,\tau \right) \right\vert ^{2}d\xi -\frac{\lambda }{6}\theta ^{2},
\end{equation}
so that the recurrence relation for the Adomian series is
\begin{equation}
\tilde{\psi}_{0}(\theta ,0)=\tilde{\Psi}\left( \theta \right) ,
\end{equation}%
\begin{eqnarray}
\tilde{\psi}_{k+1}\left( \theta ,\tau \right) &=&i\int_{0}^{\tau }\left\{
\frac{1}{2\theta }\frac{\partial ^{2}}{\partial \theta ^{2}}\left[ \theta
\tilde{\psi}_{k}(\theta ,\tau )\right] -\tilde{A}_{k}(\theta ,\tau )\right\}
d\tau ,  \notag \\
&&k=0,1,...,
\end{eqnarray}
\begin{equation}
\tilde{\Phi}_{0}(\theta ,0)=\tilde{\phi}_{1}(0)-\frac{\lambda }{6}\theta
^{2},
\end{equation}
\begin{equation}
\tilde{\Phi}_{k+1}(\theta ,\tau )=\sigma \int_{0}^{\theta }\xi \left( 1-%
\frac{\xi }{\theta }\right) \tilde{B}_{k}\left( \theta ,\tau \right) d\xi ,
\end{equation}%
where $\tilde{A}_{k}(\theta ,\tau )$ and $\tilde{B}_{k}\left( \theta ,\tau
\right) $ are the Adomian polynomials corresponding to $\tilde{\Phi}(\theta
,\tau )\tilde{\psi}(\theta ,\tau )$ and $\left\vert \tilde{\psi}\left(
\theta ,\tau \right) \right\vert ^{2}=\tilde{\psi}\left( \theta ,\tau
\right) \tilde{\psi}^{\ast }\left( \theta ,\tau \right) $, respectively.

By adopting for the initial distribution of the wave function the expression
(\ref{initial_condition}), which in the present dimensionless variables becomes $\tilde{\Psi%
}\left( \theta \right) =e^{-\theta ^{2}}$, and by assuming for the
gravitational potential a distribution at $\tau =0$ given by
\begin{equation}
\tilde{\Phi}_{0}(\theta ,0)=\tilde{a}-\frac{\lambda }{6}\theta ^{2},
\end{equation}%
where we have denoted $\tilde{\phi}_{1}(0)=\tilde{a}$, we obtain the zeroth order terms as
\begin{equation}
\tilde{A}_{0}\left( \theta ,0\right) =e^{-\theta ^{2}}\left( \tilde{a}-\frac{%
\lambda }{6}\theta ^{2}\right) ,
\end{equation}%
and
\begin{equation}
\tilde{B}_{0}\left( \xi ,0\right) =e^{-2\xi ^{2}}.
\end{equation}

Hence, we obtain the first order Adomian approximations of the wave function
and of the gravitational potential for the S-N-$\Lambda $
system as
\begin{equation*}
\tilde{\psi}_{1}(\theta ,\tau )=i\int_{0}^{\tau }\left\{ \frac{1}{2\theta }%
\frac{\partial ^{2}}{\partial \theta ^{2}}\left[ \theta \tilde{\psi}%
_{0}(\theta ,\tau )\right] -\tilde{A}_{0}(\theta ,\tau )\right\} d\tau ,
\end{equation*}%
\begin{equation}
\tilde{\psi}_{1}(\theta ,\tau )=\Bigg[\left( 2\theta ^{2}-3\right) -\left(
\tilde{a}-\frac{\lambda }{6}\theta ^{2}\right) \Bigg]e^{-\theta ^{2}}\frac{%
i\tau }{1!},
\end{equation}%
\begin{equation*}
\tilde{\Phi}_{1}(\theta ,0)=\sigma \int_{0}^{\theta }\xi \left( 1-\frac{\xi
}{\theta }\right) {\tilde{B}_{0}(\xi ,0)}d\xi ,
\end{equation*}%
\begin{equation}
\tilde{\Phi}_{1}(\theta ,0)=\frac{1}{16}\sigma \left[ 4-\frac{\sqrt{2\pi }%
\text{erf}\left( \sqrt{2}\theta \right) }{\theta }\right] .
\end{equation}
As one can see, immediately, $\tilde{\Phi}_{1}(\theta ,0)$ satisfies the
condition $\lim_{\theta \rightarrow 0}\tilde{\Phi}_{1}(\theta ,0)=0$. The
Adomian polynomial $\tilde{A}_{1}$ can be obtained immediately from Eqs.~(%
\ref{22}), and is given by
\begin{eqnarray}
\tilde{A}_{1}(\theta ,\tau ) &=&\tilde{\psi}_{0}(\theta ,0)\tilde{\Phi}{%
_{1}(\theta ,\tau )}+\tilde{\psi}{_{1}(\theta ,\theta )}\tilde{\Phi}{_{0}}%
(\theta ,0) \notag \\
&=&\frac{1}{144\theta }\Bigg\{-4\theta i\tau \left( \theta ^{2}\lambda -6%
\tilde{a}\right) 
\notag \\
&\times&[-6\tilde{a}+ \theta ^{2}(\lambda +12)-18]
\notag \\
&-&9\sqrt{2\pi }\sigma \text{erf}\left(
\sqrt{2}\theta \right) + 36\sigma \theta \Bigg\}e^{-\theta ^{2}}.
\end{eqnarray}
Thus, we obtain
\begin{equation}
\tilde{\psi}_{2}(\theta ,\tau )=i\int_{0}^{\tau }\left\{ \frac{1}{2\theta }%
\frac{\partial ^{2}}{\partial \theta ^{2}}\left[ \theta \tilde{\psi}%
_{1}(\theta ,\tau )\right] -\tilde{A}_{1}(\theta ,\tau )\right\} d\tau ,
\end{equation}%
\begin{eqnarray}
&&\tilde{\psi}_{2}(\theta ,\tau )=\left[ 4\left( \theta ^{2}-5\right) \theta
^{2}+15\right] e^{-\theta ^{2}}\frac{\left( i\tau \right) ^{2}}{2!}  \notag
\\
&+&\frac{1}{36}\left( 6\tilde{a}-\theta ^{2}\lambda \right) \left[ 6\tilde{a}%
-(\lambda +12)\theta ^{2}+18\right] e^{-\theta ^{2}}\frac{\left( i\tau
\right) ^{2}}{2!}  \notag \\
&+&\frac{1}{6}\left[ 6\tilde{a}\left( 3-2\theta ^{2}\right) +\left( 2\theta
^{4}-7\theta ^{2}+3\right) \lambda \right] e^{-\theta ^{2}}\frac{\left(
i\tau \right) ^{2}}{2!}  \notag \\
&+&\frac{\sigma \left[ \sqrt{2\pi }\text{erf}\left( \sqrt{2}\theta \right)
-4\theta \right] }{16\theta }e^{-\theta ^{2}}\frac{i\tau }{1!}.
\end{eqnarray}

The Adomian polynomial $\tilde{B}_{1}{(\theta ,\tau )}=\tilde{\psi}%
_{0}(\theta ,0)\tilde{\psi}_{1}^{\ast }(\theta ,\tau )+\tilde{\psi}%
_{1}(\theta ,\tau )\tilde{\psi}{_{0}^{\ast }(\theta ,0)}$ identically
vanishes, since $\tilde{\psi}_{0}(\theta ,0)$ is a real function, while $%
\tilde{\psi}_{1}(\theta ,\tau )$ is a purely imaginary function, so that $%
\tilde{\psi}_{1}^{\ast }(\theta ,\tau )=-\tilde{\psi}_{1}(\theta ,\tau )$.
Therefore, there is no contribution to the
gravitational potential at this order of the approximation, $\tilde{\Phi}_{2}(\theta ,\tau )=0$.
The second Adomian polynomial, defined according to $\tilde{A}_{2}{(\theta
,\tau )}=\tilde{\psi}_{0}(\theta ,0)\tilde{\Phi}{_{2}(\theta ,\tau )}+\tilde{%
\psi}_{1}(\theta ,\tau )\tilde{\Phi}_{1}(\theta ,\tau )+\tilde{\psi}%
_{2}(\theta ,\tau )\tilde{\Phi}_{0}\left( \theta ,0\right) $, is obtained as
\begin{eqnarray}
\hspace{-0.5cm}&&\tilde{A}_{2}(\theta ,\tau ) \notag \\
\hspace{-0.5cm}&=&\frac{\sigma \left[ 4\theta -\sqrt{2\pi }\text{erf}\left(
\sqrt{2}\theta \right) \right] \left[ -6\tilde{a}+\theta ^{2}(\lambda +6)-9%
\right] e^{-\theta ^{2}}}{48\theta }i\tau  \notag \\
\hspace{-0.5cm} &-&\frac{1}{432}\left( \theta ^{2}\lambda -6\tilde{a}\right) %
\Bigg\{ 36\tilde{a}^{2}-12\tilde{a}\left[ \theta ^{2}(\lambda +12)-18\right]
\notag \\
\hspace{-0.5cm}&+&\theta ^{2}(\lambda +12)\left[ \theta ^{2}(\lambda +12)-60%
\right] +18(\lambda +30)\Bigg\} e^{-\theta ^{2}}(i\tau )^{2},  \notag \\
\end{eqnarray}%
%
giving
\begin{equation}
\tilde{\psi}_{3}(\theta ,\tau )=i\int_{0}^{\tau }\left\{ \frac{1}{2\theta }%
\frac{\partial ^{2}}{\partial \theta ^{2}}\left[ \theta \tilde{\psi}%
_{2}(\theta ,\tau )\right] -\tilde{A}_{2}(\theta ,\tau )\right\} d\tau ,
\end{equation}%
and
\begin{widetext}
\begin{eqnarray}
\tilde{\psi}_{3}(\theta ,\tau ) &=&-\frac{1}{216}\Bigg\{
216\tilde{a}^{3}-108\tilde{a}^{2}\left[ \theta ^{2}(\lambda +8)-12\right] +18\tilde{a}\Bigg[ \theta
^{4}\left( \lambda ^{2}+16\lambda +48\right) -16\theta ^{2}(2\lambda
+15)+6(\lambda +30)\Bigg] \nonumber\\
&-&\theta ^{2}\lambda \left[ \theta ^{4}(\lambda
+12)^{2}-60\theta ^{2}(\lambda +12)+18(\lambda +30)\right] \Bigg\}
e^{-\theta ^{2}}\frac{(i\tau )^{3}}{3!} \nonumber\\
&+&\frac{\sigma \left[ 4\theta -\sqrt{2\pi }\text{erf}\left( \sqrt{2}\theta
\right) \right] \left[ 6\tilde{a}-(\lambda +6)\theta ^{2}+9\right] e^{-\theta ^{2}}%
}{48\theta }\frac{(i\tau )^{2}}{2!} \nonumber\\
&+&\frac{1}{432\theta }\Bigg\{ 4e^{2\theta ^{2}}\theta \Bigg[ i\tau \Bigg(
36a\left( 2\left(\tilde{a}+10\right)\theta ^{2}-3\left(\tilde{a}+5\right)-4\theta ^{4}\right) -6\lambda \left(
4\left(\tilde{a}+7\right)\theta ^{4}-(14\tilde{a}+41)\theta ^{2}+6\tilde{a}-4\theta ^{6}+9\right) \nonumber\\
&+&\theta^{2}\left( 2\theta ^{4}-11\theta ^{2}+10\right) \lambda ^{2}\Bigg) +\left(
81-54\theta ^{2}\right) \sigma \Bigg] +27\sqrt{2\pi }e^{2\theta ^{2}}\left(
2\theta ^{2}-1\right) \sigma \text{erf}\left( \sqrt{2}\theta \right)
-432\theta \sigma \Bigg\} e^{-3\theta ^{2}}\frac{(i\tau )^{2}}{2!} \nonumber\\
&+&\frac{1}{6}\left\{ \left( 4\theta ^{6}-36\theta ^{4}+75\theta
^{2}-30\right) \lambda -6\tilde{a}\left[ 4\left( \theta ^{2}-5\right) \theta ^{2}+15%
\right] \right\} e^{-\theta ^{2}}\frac{(i\tau )^{3}}{3!} \nonumber\\
&+&\frac{1}{6}\left( 8\theta ^{6}-84\theta ^{4}+210\theta ^{2}-105\right)
e^{-\theta ^{2}}\frac{(i\tau )^{3}}{3!}.
\end{eqnarray}
\end{widetext}
For the Adomian polynomial $\tilde{B}_{2}(\theta ,\tau )$, we obtain the
simple expression
\begin{equation}
\tilde{B}_{2}(\zeta ,\tau )=-\frac{1}{3}\left( 4\xi ^{2}-3\right) (\lambda
+12)e^{-2\xi ^{2}}\frac{(i\tau )^{2}}{2!},
\end{equation}%
and, thus,
\begin{eqnarray}
\tilde{\Phi}_{3}(\theta ,\tau ) &=&\sigma \int_{0}^{\theta }\xi \left( 1-%
\frac{\xi }{\theta }\right) {\tilde{B}_{2}(\xi ,\tau )}d\xi  \notag \\
&=&\frac{\sigma }{12}\left( 1-e^{-2\theta ^{2}}\right) (\lambda +12)\frac{%
(i\tau )^{2}}{2!}.
\end{eqnarray}

The next terms in the series solution of the S-N-$\Lambda$ equations
can be computed easily using the same procedure. The series expansion
simplifies significantly in the absence of the dark energy, $\lambda =0$.
In this case, the effects of self-gravitational interaction on the quantum dynamics
are described by the following approximations to the wave function:
\begin{equation}
\tilde{\psi}_{1}(\theta ,\tau )=-\left( \tilde{a}-2\theta ^{2}+3\right)
e^{-\theta ^{2}}\frac{i\tau }{1!},
\end{equation}%
\begin{eqnarray}
\tilde{\psi}_{2}(\theta ,\tau ) &=&\Bigg\{\left[ 4\left( \theta
^{2}-5\right) \theta ^{2}+15\right]  \notag \\
&+&\tilde{a}\left( 3-2\theta ^{2}\right) +\tilde{a}\left( \tilde{a}-2\theta
^{2}+3\right) \Bigg\}e^{-\theta ^{2}}\frac{(i\tau )^{2}}{2!}  \notag \\
&+&\frac{\sigma \left[ \sqrt{2\pi }\text{erf}\left( \sqrt{2}\theta \right)
-4\theta \right] e^{-\theta ^{2}}}{16\theta }\frac{i\tau }{1!},
\end{eqnarray}
\begin{eqnarray}
&&\tilde{\psi}_{3}(\theta ,\tau )=\left( 8\theta ^{6}-84\theta
^{4}+210\theta ^{2}-105\right) e^{-\theta ^{2}}\frac{(i\tau )^{3}}{3!}
\notag \\
&+&\tilde{a}e^{-\theta ^{2}}\left[ 4\left( \theta ^{2}-5\right) \theta ^{2}+15%
\right] \frac{(i\tau )^{3}}{3!}  \notag \\
&-&\tilde{a}e^{-\theta ^{2}}\left[ \tilde{a}^{2}+\tilde{a}\left( 6-4\theta
^{2}\right) +4\theta ^{4}-20\theta ^{2}+15\right] \frac{(i\tau )^{3}}{3!}
\notag \\
&+&\frac{\sigma \left( 2\tilde{a}-2\theta ^{2}+3\right) \left[ 4\theta -\sqrt{%
2\pi }\text{erf}\left( \sqrt{2}\theta \right) \right] e^{-\theta ^{2}}}{%
16\theta }\frac{(i\tau )^{2}}{2!}  \notag \\
&+&\frac{1}{48\theta }\Bigg\{e^{2\theta ^{2}}\Bigg[4\theta \Bigg(4\tilde{a}%
i\tau \Bigg(2(\tilde{a}+10)\theta ^{2}-3(\tilde{a}+5)-4\theta ^{4}\Bigg)
\notag \\
&+&\left( 9-6\theta ^{2}\right) \sigma \Bigg)+3\sqrt{2\pi }\left( 2\theta
^{2}-1\right) \sigma \text{erf}\left( \sqrt{2}\theta \right) \Bigg]  \notag
\\
&-&48\theta \sigma \Bigg\}e^{-3\theta ^{2}}\frac{(i\tau )^{2}}{2!}.
\end{eqnarray}

We again represent the probability density $P\left( \theta ,\tau ,\sigma ,\lambda
\right) =\left\vert \tilde{\psi}_{3}(\theta ,\tau )\tilde{\psi}_{3}^{\ast
}(\theta ,\tau )\right\vert ^{2}$ by its Pad\'{e} approximant $P[m/n]\left( \theta ,\tau ,\sigma
,\lambda \right) $. In the first order of approximation, we obtain
\begin{equation}
P^{(1)}[1/2]\left( \theta ,\tau ,\sigma ,\lambda \right) \simeq \frac{%
e^{-2\theta ^{2}}}{1-\left[ 6\tilde{a}-(\lambda +12)\theta ^{2}+18\right]
^{2}\tau ^{2}/36}.
\end{equation}
and at the second order of approximation, 
\begin{eqnarray}
\hspace{-0.3cm}&&P^{(2)}[1/2]\left( \theta ,\tau ,\sigma ,\lambda \right)
\simeq e^{-2\theta ^{2}}  \notag \\
\hspace{-0.3cm}&\times&\Bigg\{ 1- \frac{\sqrt{\frac{\pi }{2}}\sigma \tau ^{2}\text{%
erf}\left( \sqrt{2}\theta \right) \left[ -12a-3\sigma -36+2\left( \lambda
+12\right) \theta ^{2}\right] }{48\theta }  \notag \\
\hspace{-0.3cm}&+&\frac{1}{48}\Bigg[ 288+24\lambda -3\sigma ^{2}-72\sigma
-24a\sigma +\Big( 4\lambda \sigma -32\lambda \notag \\
\hspace{-0.3cm}&+&48\sigma -384\Big) \theta ^{2}\Bigg] \tau ^{2}-\frac{\pi
\sigma ^{2}\text{erf} ^{2}\left( \sqrt{2}\theta \right)}{128\theta ^{2}}\tau
^{2}\Bigg\} ^{-1}.
\end{eqnarray}
To third order, the probability density can be
approximated as 
\begin{eqnarray}
&&P^{(3)}[1/2]\left( \theta ,\tau ,\sigma ,\lambda \right) \simeq
e^{-2\theta ^{2}}\Bigg\{1+\Bigg[-\frac{\pi \sigma ^{2}\text{erf}\left( \sqrt{%
2}\theta \right) ^{2}}{128\theta ^{2}}  \notag \\
&+&\frac{\sqrt{\frac{\pi }{2}}\sigma ^{2}\text{erf}\left( \sqrt{2}\theta
\right) }{16\theta }+\frac{\sqrt{\frac{\pi }{2}}\sigma \text{erf}\left(
\sqrt{2}\theta \right) }{4\theta }-\frac{2\theta ^{2}\lambda }{3}%
-e^{-2\theta ^{2}}\sigma   \notag \\
&-&8\theta ^{2}+\frac{\lambda }{2}-\frac{\sigma ^{2}}{16}+6\Bigg]\tau ^{2}%
\Bigg\}^{-1}.
\end{eqnarray}%
%
%
Higher order approximations of the probability density of a Gaussian wave
packet, evolving under self-gravity in the presence of dark energy,
can also be calculated easily with the aid of computer algebra systems.
The gravitational self-potential can be approximated as
\begin{eqnarray}
\tilde{\Phi}(\theta, \tau)&=&\tilde{a}+\frac{1}{16} \sigma \left[4-\frac{%
\sqrt{2 \pi } \text{erf}\left(\sqrt{2} \theta \right)}{\theta }\right]
\notag \\
&-& \frac{1}{24} \left(1-e^{-2 \theta ^2}\right) (\lambda +12) \sigma \tau ^2-%
\frac{\theta ^2 \lambda }{6},
\end{eqnarray}
or, in terms of the Pad\'{e} approximants of the power series,
\begin{widetext}
\bea
\tilde{\Phi}[1/2](\theta, \tau)=-\frac{\left[48 a \theta -3 \sqrt{2 \pi } \sigma  \text{erf}\left(\sqrt{2} \theta \right)-8
   \theta ^3 \lambda +12 \theta  \sigma \right]^2}{48 \theta  \left\{-48 a \theta +3 \sqrt{2
   \pi } \sigma  \text{erf}\left(\sqrt{2} \theta \right)+8 \theta ^3 \lambda +2 \theta
   \sigma  \left[\left(e^{-2 \theta ^2}-1\right) (\lambda +12) \tau ^2-6\right]\right\}}.
\eea
\end{widetext}

\subsection{The dark energy dominated regime}

We now consider the limiting case in which the dark energy density dominates
the matter density, $\lambda \gg \sigma \left\vert \tilde{\psi}%
(\theta ,\tau )\right\vert $. The Poisson equation then takes the simple
form
\begin{equation}
\frac{\partial ^{2}}{\partial \theta ^{2}}\left[ \theta \tilde{\Phi}(\theta
,\tau )\right] =-\lambda \theta ,
\end{equation}%
and can be immediately integrated to give
\begin{equation}
\tilde{\Phi}(\theta )=\tilde{a}-\frac{\lambda \theta ^{2}}{6},
\end{equation}%
where we have assumed that the background gravitational potential is
independent of time. Hence, for the dark energy dominated phase, the
Schr{\"o}dinger equation takes the form
\begin{equation}
\frac{\partial \tilde{\psi}(\theta ,\tau )}{\partial \tau }=i\left\{ \frac{1%
}{2\theta }\frac{\partial ^{2}}{\partial \theta ^{2}}\left[ \theta \tilde{%
\psi}(\theta ,\tau )\right] -\left( \tilde{a}-\frac{\lambda \theta ^{2}}{6}%
\right) \tilde{\psi}(\theta ,\tau )\right\} ,
\end{equation}%
and can be formally solved to give
\bea
\hspace{-0.9cm}&&\tilde{\psi}(\theta ,\tau )=\tilde{\Psi}\left( \theta \right) + \nonumber\\
\hspace{-0.9cm}&&i\hat{L}%
_{\tau }^{-1}\left\{ \frac{1}{2\theta }\frac{\partial ^{2}}{\partial \theta
^{2}}\left[ \theta \tilde{\psi}(\theta ,\tau )\right] -\left( \tilde{a}-%
\frac{\lambda \theta ^{2}}{6}\right) \tilde{\psi}(\theta ,\tau )\right\} .
\eea

By decomposing the wave function as  $\tilde{\psi}(\theta ,\tau
)=\sum_{n=0}^{\infty }$ $\tilde{\psi}_{n}(\theta ,\tau )$, we obtain the
following recurrence relations for the determination of the components  $%
\tilde{\psi}_{n}(\theta ,\tau )$:
\begin{equation}
\tilde{\psi}_{0}(\theta ,\tau )=\tilde{\Psi}\left( \theta \right) ,
\end{equation}
\begin{eqnarray}
\hspace{-0.5cm}&&\tilde{\psi}_{k+1}(\theta ,\tau ) =\nonumber\\
\hspace{-0.5cm}&&i\int_{0}^{\tau }\left\{ \frac{1}{%
2\theta }\frac{\partial ^{2}}{\partial \theta ^{2}}\left[ \theta \tilde{\psi}%
_{k}(\theta ,\xi )\right] -\left( \tilde{a}-\frac{\lambda \theta ^{2}}{6}%
\right) \tilde{\psi}_{k}(\theta ,\xi )\right\} d\zeta , \nonumber\\
\hspace{-0.5cm}&&k =0,1,2,...,n,
\end{eqnarray}
and the first few approximations of the quantum wave packet in the dark energy dominated regime are obtained as
\be
\tilde{\psi}_0=e^{-\theta ^2},
\ee
\be
\tilde{\psi}_1(\theta, \tau )=\frac{1}{6} e^{-\theta ^2}  \left[6 \tilde{a}
   -(\lambda -12)\theta ^2-18\right]\frac{i \tau}{1!},
\ee
\bea
\hspace{-0.3cm}&&\tilde{\psi}_2(\theta, \tau)=\frac{1}{36} e^{-\theta ^2}   \Bigg\{(36 a^2-12
   a \left[\theta ^2 (\lambda
   -12)+18\right] \nonumber\\
 \hspace{-0.3cm} &+& \theta ^4
   (\lambda -12)^2+60 \theta
   ^2 (\lambda -12)-18
   (\lambda -30)\Bigg\}\frac{\left(i\tau\right)^2}{2!}, \nonumber\\
\eea
\bea
\hspace{-0.2cm}&&\tilde{\psi}_3(\theta, \tau)= \frac{1}{216}\Bigg\{e^{-\theta ^2} \Bigg[216 a^3-108 a^2
   \left(\theta ^2 (\lambda
   -12)+18\right)  \nonumber\\
 \hspace{-0.2cm} &+& 18 a
   \left(\theta ^4 (\lambda
   -12)^2+60 \theta ^2
   (\lambda -12)-18 (\lambda
   -30)\right)  \nonumber\\
 \hspace{-0.2cm} &-& \theta ^6
   (\lambda
   -12)^3-126 \theta ^4
   (\lambda -12)^2  \nonumber\\
 \hspace{-0.2cm} &+& 6 \theta ^2
   \left(13 \lambda ^2-786
   \lambda +7560\right)+108
   (13 \lambda
   -210)\Bigg]\Bigg\}\frac{(i \tau)
   ^3}{3!}. \nonumber\\
\eea
The probability density of the Gaussian wave packet can be obtained, with the help of the Pad\'{e} approximants, to different orders of approximation, as
\be
\tilde{P}[1/2](\theta, \tau)\simeq \frac{6 e^{-2 \theta    ^2}}{\left(4 \theta    ^2-3\right) (\lambda -12)   \tau ^2+6},
   \ee
   and
   \begin{widetext}
   \be
   \tilde{P}[2/3]\simeq \frac{e^{-2 \theta ^2}   \left\{\tau ^2 \left[-8   \theta ^2 \left(6 \theta ^2   (\lambda -12)-5 \lambda   +36\right)-21 \lambda
   +108\right]+36 \left(4   \theta   ^2-3\right)\right\}}{\tau ^2   \left\{8 \theta ^2 \left[6   \theta ^2 (\lambda -12)-13   \lambda +180\right]+33
   \lambda -540\right\}+36   \left(4 \theta ^2-3\right)},
   \ee
   \end{widetext}
   and so on. The analytical expressions for the probability density can also be obtained easily to any desired order of approximation.

\subsection{Numerical analysis}

\begin{figure}[h!]
\centering
\includegraphics[width=7cm]{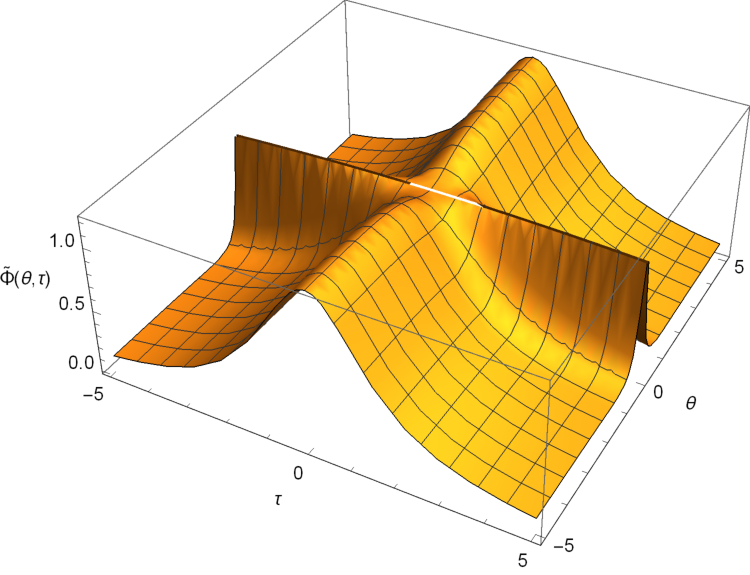}
\caption{Variation of the rescaled self-gravitational potential $\tilde{\Phi}(%
\theta, \tau)$ in the absence of dark energy ($\lambda %
=0$), for $\sigma =1$. The initial value of the potential is chosen as $\tilde{a}=1$. }
\label{fig2}
\end{figure}

In the final part of this section, we consider the numerical results obtained from the Adomian series
solutions of the S-N-$\Lambda$ system. Our main goal
is to highlight the effects of the self-gravitational potential and the dark
energy density on the evolution of the probability density associated with the Gaussian quantum wave packet.
In Fig.~\ref{fig2}, we present the
three-dimensional evolution of the rescaled gravitational potential in the
absence of dark energy, i.e., with $\lambda =0$, and with $\sigma =1$.
For convenience, we take $\tilde{a}=1$ as its initial value.

In this case, the gravitational potential can be approximated  by
\be
\tilde{\Phi}(\theta,\tau)\simeq\frac{4 \theta  (4 \tilde{a}+\sigma )-\sqrt{2 \pi } \sigma  \text{erf}\left(\sqrt{2} \theta \right)}{16 \theta
   \left[\frac{8 e^{-2 \theta ^2} \left(e^{2 \theta ^2}-1\right) \theta  \sigma  \tau ^2}{4 \theta  (4
   \tilde{a}+\sigma )-\sqrt{2 \pi } \sigma  \text{erf}\left(\sqrt{2} \theta \right)}+1\right]},
\ee
and the full solution satisfies the condition $\lim_{\tau \rightarrow \infty}\tilde{\Phi}(\theta,\tau) =0$. Mathematically, a singularity develops in $\tilde{\Phi}(\theta,\tau)$ for values of $\theta $ satisfying $4 \theta  (4 a+\sigma)-\sqrt{2 \pi } \sigma \text{erf}\left(\sqrt{2}\theta \right) = 0$. However, to at least third order in the approximation, this equation does not have any real roots, except at $\theta =0$.

\begin{figure*}[tbp!]
\centering
\includegraphics[width=7cm]{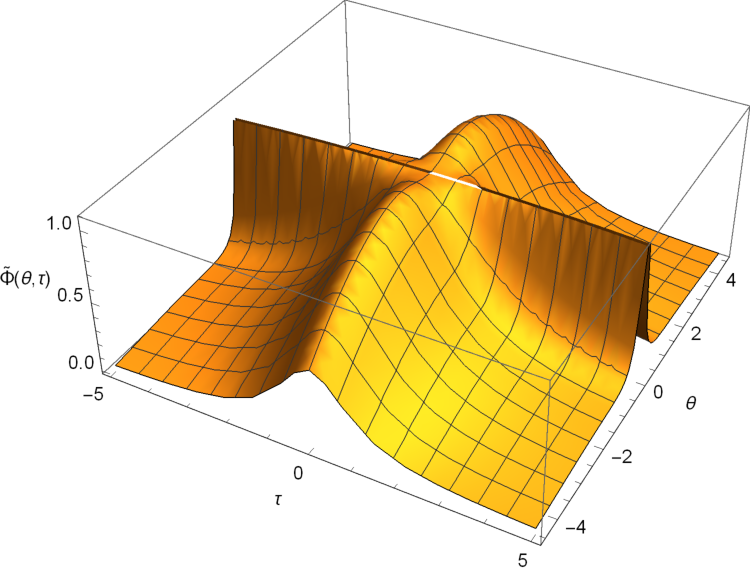}
\includegraphics[width=7cm]{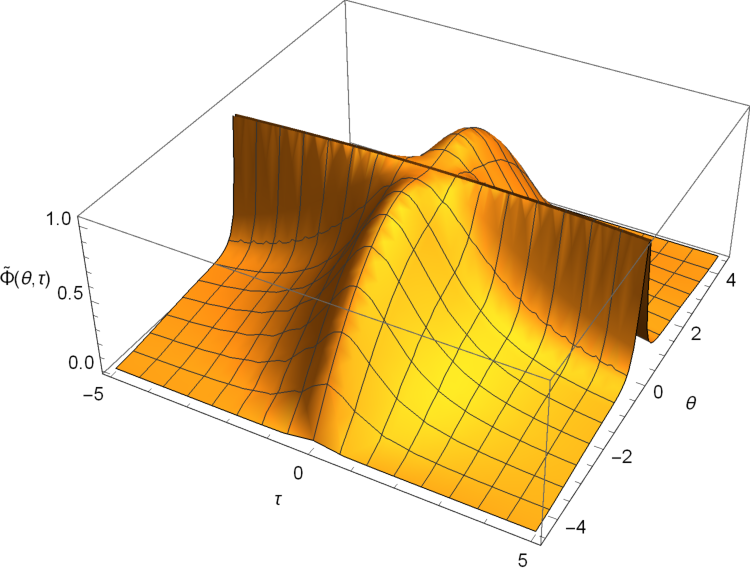}
\caption{Variation of the gravitational potential $\tilde{\Phi}(%
\theta, \tau)$ for $\lambda =0.20$ (left panel), and for $%
\lambda =0.35$ (right panel), for $\sigma =1$. For the
initial value of the potential we have adopted the value $\tilde{a}=1$.}
\label{fig3}
\end{figure*}

The variation of the self-gravity potential in the presence of dark energy
is represented in Fig.~\ref{fig3}, for two different values of $\lambda$; $%
\lambda =0.20$ and $\lambda =0.35$. In the large $\theta $ limit its behavior can be approximated as
\be
\tilde{\Phi}(\theta,\tau)\simeq \frac{12\tilde{ a}-2 \lambda \theta ^2
   +3 \sigma }{12
   \left\{\frac{
   (\lambda  +12
   )\sigma\tau ^2}{2 \left[12 \tilde{a}a-2 \theta ^2
   \lambda +3 \sigma
   \right]}+1\right\}}.
\ee
Thus, we see that the presence of a positive cosmological constant does have a significant effect on the distribution of the gravitational potential.
To at least the considered order of approximation, the condition $\lim_{\tau \rightarrow \infty}\tilde{\Phi}(\theta,\tau) =0$ still holds. On the other hand, as expected, $\lim_{\lambda \rightarrow \infty}\tilde{\Phi}(\theta,\tau) =-\infty$. In both cases, the $\tilde{\Phi}(\theta,\tau)$ has a sharp maximum at the origin of the coordinate system, $\theta =0$.

\begin{figure*}[tbp!]
\centering
\includegraphics[width=7cm]{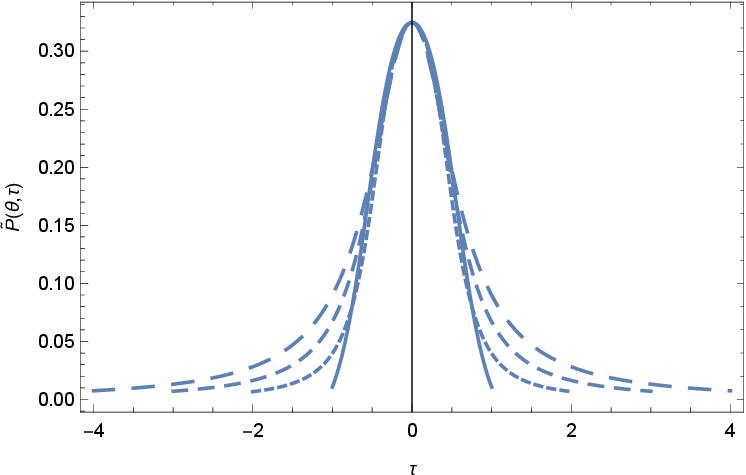}
\includegraphics[width=7cm]{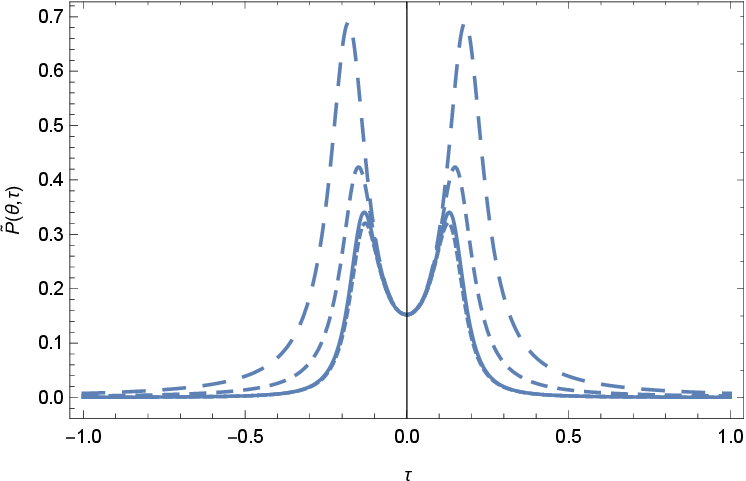}
\caption{Variation of the probability density $\tilde{P}(\theta,%
\tau)$ at fixed values of $\theta $. In the left-hand panel $%
\theta =0.75$, while $\tilde{a}=0$, $\sigma =0$, $%
\lambda =0$ (solid curve), $\tilde{a}=1$, $\sigma =1$, $%
\lambda =0$ (dotted curve), $\tilde{a}=1$, $\sigma =1$, $%
\lambda =5$ (dotted curve), and $\tilde{a}=1$, $\sigma =1$, and $%
\lambda =7$ (long dashed curve). In the right-hand panel $%
\theta =3$, while $\tilde{a}=0$, $\sigma =0$, $%
\lambda =0$ (solid curve), $\tilde{a}=1$, $\sigma =1$, $%
\lambda =0$ (dotted curve), $\tilde{a}=1$, $\sigma =1$, $
\lambda =1$ (dotted curve), and $\tilde{a}=1$, $\sigma =1$, and $%
\lambda =1.5$ (long dashed curve). For the sake of
presentation the probability density was multiplied by a factor of $10^7$. }
\label{fig4}
\end{figure*}

The time variation of the probability density of the Gaussian wave packet is
represented, for fixed values of the radial coordinate $\theta$, in Figs.~%
\ref{fig4}. There are two significant effects induced by the presence of the dark energy.
As one can see from the left-hand panel, for (relatively) small values of the dimensionless radial coordinate $\theta$, the probability density in the presence of $\Lambda > 0$ almost coincides with the function describing the evolution with $\Lambda=0$, for $-1\leq \tau \leq 1$, and has the same maximum value. In the absence of $\Lambda$, the probability density tends to zero at a finite value of $\tau$. However, dark energy significantly modifies the tail of the Gaussian distribution, which extends in time and induces much higher values of the probability density, as compared to the $\Lambda =0$ case. From the right-hand panel we see that, for larger values of $\theta$, the dark energy has two different effects on the probability density. The first is a significant increase in the amplitude of the probability density, with the maximum increased by a factor of at least two. This indicates the increased probability of finding the wave packet at larger distances from the center, the effect being a direct consequence of the presence of repulsive dark energy. Secondly, at large distances, the probability density tends to zero. However, the decrease is much slower for $\Lambda >0$, and is directly correlated with the increase of the amplitude of the wave. Another interesting effect is related to the change in the shape of the wave function function, which evolves from a single-peaked into a double-peaked symmetric function.

The three-dimensional evolution of the wave packet in the presence
of the self-gravitational field and the dark energy density is depicted in Figs.~\ref%
{fig5}. The same effects, as previously mentioned, are also apparent when considering the three-dimensional evolution of the wave packet. For large values of $\tau$ and $\theta$, $P(\theta,\tau)\rightarrow 0$, but the dynamics of the transition to the asymptotic limit are strongly influenced by the presence of dark energy, whose effect becomes significant at late times and for large values of the radial coordinate.

\begin{figure*}[tbp!]
\centering
\includegraphics[width=7cm]{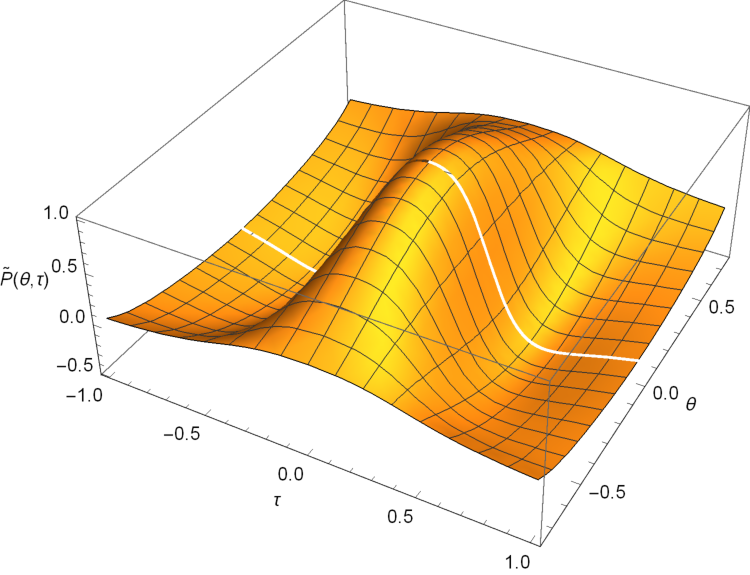}
\includegraphics[width=7cm]{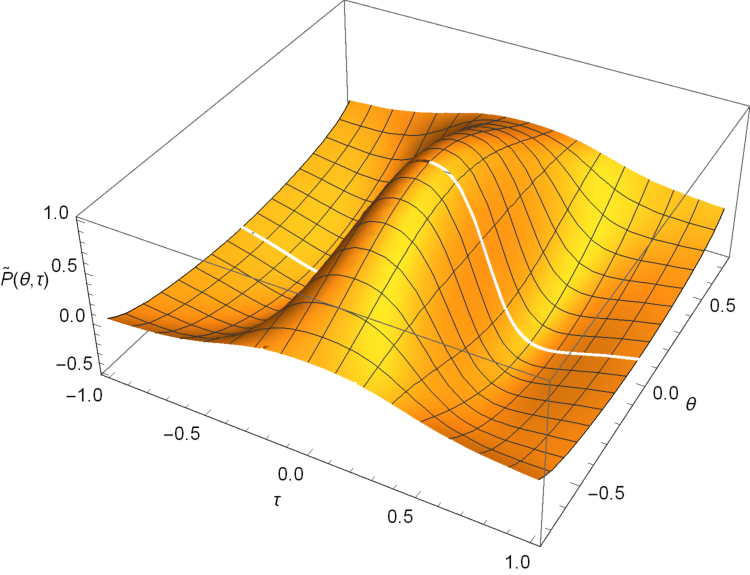}
\caption{Three-dimensional variation of the probability density $\tilde{P}(%
\theta,\tau)$ for $\tilde{a}=1$, $\sigma =1$, and $%
\lambda =0$ (left panel), and for $\tilde{a}=1$, $\sigma =1$%
, and $\lambda =1$ (right panel), respectively.}
\label{fig5}
\end{figure*}

\begin{figure*}[tbp!]
\centering
\includegraphics[width=7cm]{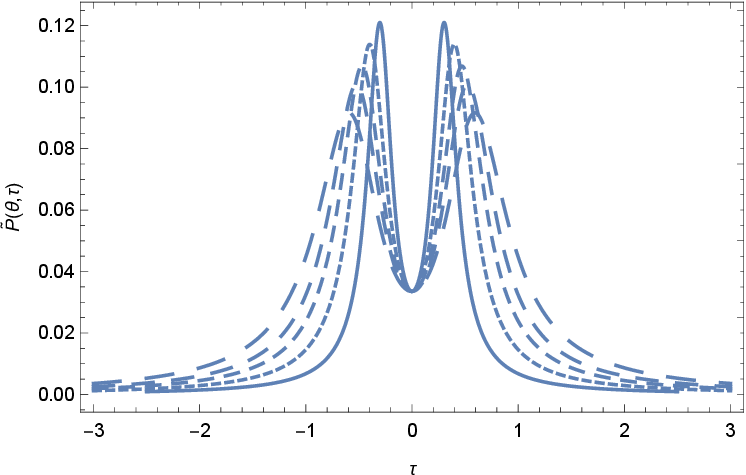}
\includegraphics[width=7cm]{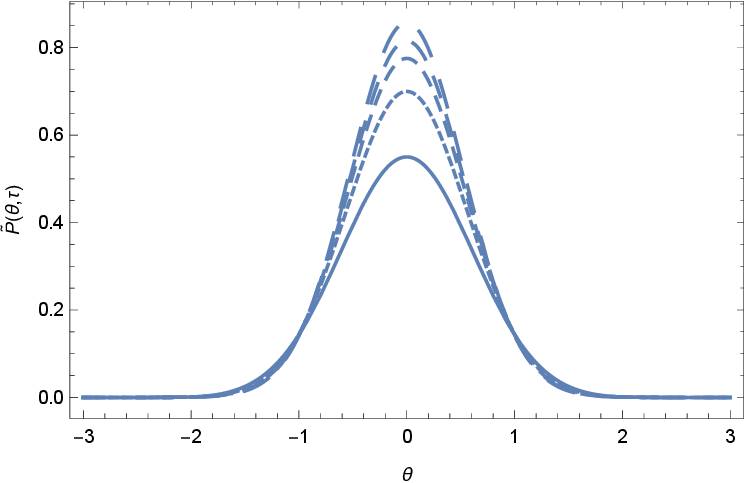}
\caption{Variation of the probability density $\tilde{P}(\theta, \tau)$ in the dark energy dominated regime, for a fixed value of the radial coordinate, $\theta =2$ (left-hand panel), and for a fixed time, $\tau =0.35$ (right-hand panel). In each case, we plot the probability density for a range of parameter values, namely, $\tilde{a}=0$, $\lambda =0$, which corresponds to the free evolution of the Gaussian wave packet (solid curve), $\tilde{a}=1$, $\lambda =5$ (dotted curve), $\tilde{a}=1$, $\lambda =7$ (short dashed curve), $\tilde{a}=1$, $\lambda =8$ (dashed curve), and $\tilde{a}=1$, $\lambda =9$ (long dashed curve).}
\label{fig6}
\end{figure*}

The behavior of the probability density in the dark energy dominated regime is presented in Fig.~\ref{fig6}, for two distinct physical situations, corresponding to a fixed value of $\theta$ (left panel), and to a fixed value of $\tau$ (right panel). Even though, in this regime, there is a qualitative similarity with the $\Lambda=0$ case, significant differences also appear. The double-peaked shape of the Gaussian distribution is extended in time for fixed $\theta$, and the shape of the Gaussian tail is strongly modified, indicating an increase in the probability of finding the particle at higher values of $\tau$. Moreover, the maximum value of the probability as a function of $\theta$, at a given time, increases dramatically with increasing $\lambda$. However, at least at the considered order of approximation, and for the adopted values of $\lambda$ in the large $\theta$ limit, the probability distribution of the initially Gaussian wave packet still tends to zero. Nonetheless, much larger values of $\lambda$, in the range $\lambda \in \left\{10^2,10^3\right\}$, would greatly modify the dynamics of the wave packet at infinity.

\section{Comparison of the Adomian method with previous analytical and numerical results} \label{sect5}

A particle obeying the S-N-$\Lambda$ equation of motion experiences three tendencies in its dynamics.
Both canonical quantum diffusion and dark energy induced acceleration cause its wave function to spread whereas Newtonian self-gravity, represented by the non-linear term, acts to localize the wave packet.
In \cite{Matt}, it was argued that the relative strengths of these three tendencies can be estimated, at least approximately, by considering the motion of the peak radial probability density, $r_{p}(t)$.
This is the position of the spherical shell at which the radial probability density $dP/dr = 4\pi r^2 |\psi|^2$ reaches its maximum, that is, the radius at which the particle is most likely to be found at a given time $t$.
It is determined by solving the equation
\begin{eqnarray} \label{r_p(t)}
\left|\psi(r,t)\right|^2 + r \left|\psi(r,t)\right| \frac{d\left|\psi(r,t)\right|}{dr} = 0 \, ,
\end{eqnarray}
or, alternatively,
\be\label{r_p(theta)}
\left|\tilde{\psi}\left(\theta,\tau\right)\right|^2+\theta \left|\tilde{\psi}\left(\theta,\tau\right)\right|\frac{d\left|\tilde{\psi}\left(\theta,\tau\right)\right|}{d\theta}=0,
\ee
which is equivalent to setting $d^2P(r,t)/dr^2 = 0$ or $d^2\tilde{P}(\theta,\tau)/d \theta ^2=0$, respectively.

The contributions to the total acceleration experienced by $r_{p}(t)$ due to canonical quantum diffusion, self-gravity, and dark energy are then estimated as
\begin{eqnarray} \label{a_SE}
a_{\mathrm{SE}} \simeq \frac{\hbar^2}{m^2 r_p^3} \, ,
\end{eqnarray}
\begin{eqnarray} \label{a_SN}
a_{\mathrm{SN}} \simeq -\frac{Gm}{r_{p}^{2}} \, ,
\end{eqnarray}
and
\begin{eqnarray} \label{a_Lambda}
a_{\Lambda} \simeq \frac{\Lambda c^{2}}{3} r_{p} \, .
\end{eqnarray}
The subscript SE refers to the canonical Schr{\" o}dinger equation, SN refers to the standard Schr{\" o}dinger-Newton contribution, and $\Lambda$ denotes the additional term induced by the dark energy density.

In order to determine the regimes in which the different tendencies dominate the dynamics, we consider equality between the absolute magnitudes of the accelerations (\ref{a_SE})-(\ref{a_Lambda}) in a pair-wise manner, i.e.,
\begin{eqnarray} \label{pair-wise-1}
a_{\mathrm{SE}} = |a_{\mathrm{SN}}| \, , \quad r_{p}^{(1)} &\simeq& \frac{\lambda_{\rm C}^3(m)}{l_{\rm Pl}^2} \simeq \frac{l_{\rm Pl}^4}{r_{\rm S}^3(m)},
\end{eqnarray}
or, equivalently,
\be
r_p^{(1)}\simeq 3.563\times 10^{24}\times \left(\frac{m}{m_p}\right)^{-3}\;{\rm cm} ,
\ee
plus
\begin{eqnarray} \label{pair-wise-2}
a_{\mathrm{SE}} = a_{\mathrm{\Lambda}} \, , \quad r_{p}^{(2)}& \simeq& \sqrt{\lambda_{\rm C}(m)l_{\rm dS}} \, ,
\end{eqnarray}
giving
\be
r_{p}^{(2)} \simeq 1.907\times 10^7\times \left(\frac{m}{m_p}\right)^{-1/2}\left(\frac{\Lambda}{10^{-56}\;{\rm cm^{-2}}}\right)^{-1/4}\;{\rm cm} ,
\ee
and
\begin{eqnarray} \label{pair-wise-3}
|a_{\mathrm{SN}}| = a_{\mathrm{\Lambda}} \, , \quad r_{p}^{(3)} &\simeq& (r_{\rm S}(m)l_{\rm dS}^2)^{1/3} \, ,
\end{eqnarray}
yielding
\be
r_{p}^{(3)}\simeq 42.0491\times \left(\frac{m}{m_p}\right)^{1/3}\left(\frac{\Lambda}{10^{-56}\;{\rm cm^{-2}}}\right)^{-1/3}\;{\rm cm},
\ee
where $\lambda_{\rm C}(m) = \hbar/(mc)$ is the reduced Compton wavelength of the particle, $r_{\rm S}(m) = 2Gm/c^2$ is its Schwarzschild radius, $l_{\rm Pl} = \sqrt{G\hbar/c^3} \simeq 10^{-33}$ cm is the Planck length, and $l_{\rm dS} = \sqrt{3/\Lambda} \simeq 10^{28}$ cm is de Sitter radius.
Note that the latter is comparable to the present day radius of the Universe \cite{Hobson:2006se} and that we have neglected numerical factors of order unity in all three equations.

The critical value of $r_p(t)$ in Eq. (\ref{pair-wise-3}) is the classical turn-around radius for a spherical compact object in the Schwarzschild-de Sitter spacetime \cite{r_TU},
\begin{eqnarray} \label{r_TU}
r_{\leftrightarrow} = \left(\frac{3Gm}{\Lambda c^2}\right)^{1/3} \, ,
\end{eqnarray}
but the critical values given in Eqs. (\ref{pair-wise-1}) and (\ref{pair-wise-2}) include genuine quantum effects.
The absolute magnitudes of all three contributions are equal when
\begin{eqnarray} \label{Compton_critical}
\lambda_{\rm C}(m) \simeq (l_{\rm Pl}^4l_{\rm dS})^{1/5} \simeq 10^{-21} \, {\rm cm} \, ,
\end{eqnarray}
or, equivalently,
\begin{eqnarray} \label{m_critical}
m \simeq (m_{\rm Pl}^4m_{\rm dS})^{1/5} \simeq 10^{-17} \, {\rm g} \, ,
\end{eqnarray}
where $m_{\rm Pl} = \sqrt{\hbar c/G} \simeq 10^{-5}$ g and $m_{\rm dS} = (\hbar/c)\sqrt{\Lambda/3} \simeq 10^{-66}$ g are the Planck mass and the de Sitter mass, respectively.
The approximate value of the peak radial probability is
\begin{eqnarray} \label{r_p_critical}
r_p \simeq  (l_{\rm Pl}^2l_{\rm dS}^3)^{1/5} \simeq 10^4 \, {\rm cm} \, ,
\end{eqnarray}
and a more careful estimate, accounting accurately for numerical factors, gives $r_p \simeq 67$ m, as shown in \cite{Matt}.

For a Gaussian distribution, the initial peak radial probability is comparable to the initial width of the wave function, $r_p(0) \simeq \sigma_0$, and the two are equivalent up to a multiplicative constant of order one for a large class of physically reasonable wave functions \cite{Matt}.
Therefore, Eq. (\ref{r_p_critical}) also gives the order of magnitude value of the minimum initial width required, in order for the acceleration due to dark energy to dominate both canonical quantum diffusion and self-gravitation.

This is a very clear and somewhat surprising prediction: in the S-N-$\Lambda$ system, the spreading of any spherically symmetric wave packet with an initial width $\sigma_0 \gtrsim 67$ m will be dominated by the accelerated expansion of the Universe, due to dark energy, regardless of its initial mass.
For particles with masses $m \gtrsim 10^{-17}$ g, Eq. (\ref{pair-wise-2}) implies that the dark energy term always dominates over canonical quantum diffusion, whenever the initial width of the wave packet exceeds this critical value.
For heavier particles, we expect the outer shells of the wave packet to undergo accelerated expansion due to dark energy while the inner core region contracts under self-gravity.
By Eq. (\ref{pair-wise-3}), the critical radius marking the division between collapsing and expanding shells should be of the order of the classical turn-around radius (\ref{r_TU}).

However, these very strong predictions were derived using rather crude analytical techniques and approximations.
It is therefore reasonable to ask: can they be trusted?
To answer this question, the numerical solution of the S-N-$\Lambda$ system was presented in \cite{Matt}, for an initially Gaussian wave packet with a range of initial widths and particle masses.
Remarkably, the existence of both a critical mass of order $10^{-17}$ g, and of critical initial width of order $\sigma_0 \simeq 6.7 \times 10^2$ cm, was verified by the numerical results.
A summary of the numerical results obtained in \cite{Matt}, for a particle wave function of initial width $\sigma_0 = 7.5 \times 10^2$ cm, and particle masses in the range $10^{-18} \, {\rm kg} \leq m \leq 10^{-16} \, {\rm kg}$, is given in Table 1.

\setlength{\tabcolsep}{6pt}
\renewcommand{\arraystretch}{1.7}
\begin{widetext}
\begin{center}
\begin{table}[!t]
\label{tab: SNLambda results}
    \begin{tabular}{|p{4cm}|p{3.8cm}|}
        \hline
        Mass                         & Behavior                                \\ \hline \hline
        Below $1 \times 10^{-18}$ g                                       & Evolution indistinguishable from that of a free particle in canonical quantum mechanics \\ \hline
        $2\times 10^{-18}$ g to $3\times 10^{-17}$ g             & The whole wave packet spreads faster than that of a canonical free particle \\ \hline
        $4 \times 10^{-17}$ g to $5 \times 10^{-17}$ g           & The inner core of the wave function spreads slower than the wave function of the canonical free particle while the outer shells spread faster \\ \hline
        $6\times 10^{-17}$ g to $1 \times 10^{-16}$ g            & The inner core of the wave function collapses under self-gravity while the outer shells spread faster than in canonical quantum mechanics  \\ \hline
	$ \sim 2 \times 10^{-16}$ g                                         & Chaotic \\ \hline
        Above $3 \times 10^{-16}$ g                                       & Stationary \\ \hline
    \end{tabular}
    \caption{Dynamical evolution of a Gaussian wave packet, with initial width $\sigma_0 = 7.5 \times 10^2$ cm, under the S-N-$\Lambda$ equation, according to the numerical solution obtained in \cite{Matt}.
The comparison is made to a free particle in canonical quantum mechanics, evolving under the canonical Schr{\" o}dinger equation.
}
\end{table}
\end{center}
\end{widetext}

We note that the chaotic and stationary regimes obtained for larger values of $m$ are artifacts of the numerics, which were unable to probe masses above $\sim 2 \times 10^{-16}$ g due to limited computational resources.
The critical radius marking the boundary between the collapsing inner core and the expanding outer shells of the wave packet was also verified to be within one order of magnitude of the classical turn-around radius (\ref{r_TU}), which isn't bad for such a crude analysis \cite{Matt}.

We now demonstrate, explicitly, that the series solution of the S-N-$\Lambda$ system, obtained using the ADM, is consistent with previous numerical results and analytical estimates.
To do this, we first estimate $\theta _p^{(0)}(\tau)$ for the free Gaussian wave packet, as given by Eq.~(\ref{thetap0}), which can be obtained as
\be
\theta _p^{(0)}(\tau)=\frac{1}{2}\sqrt{1+4\tau^2}.
\ee
To first order in the approximation, and keeping only the background gravitational potential and the dark energy terms in the series expansion, we obtain
\begin{widetext}
\bea
&&\theta _p^{(1)}(\tau)\simeq \frac{1}{2 \sqrt{3}}\times \nonumber\\
&&\sqrt{\frac{24 (\lambda +12)   \tau ^6 \left(8 \tilde{a}+\lambda +36\right)+3   (\lambda +12) \tau ^4 (32 \tilde{a}+5   \lambda +156)+2 \tau ^2 (6
   \tilde{a}+\lambda +30) (12 \tilde{a}+\lambda   +48)-16 (\lambda +12)^2 \tau   ^8+144}{\tau ^2 \left(4 (5 \tilde{a}+21)   \lambda +48 (\tilde{a}
   (\tilde{a}+11)+27)+\lambda  ^2\right)+48}}.\nonumber\\
\eea
\end{widetext}

\begin{figure*}[tbp!]
\centering
\includegraphics[width=7cm]{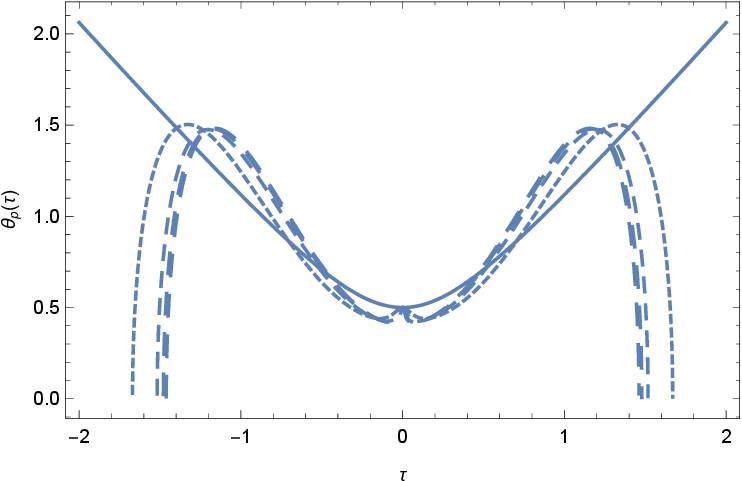}
\caption{Variation of  $\theta _p(\tau)$ as a function of the dimensionless time $\tau$, for $\tilde{a} =1$, and for different values of $\lambda$: $\lambda =50$ (dotted curve), $\lambda =150$ (short dashed curve), $\lambda =250$ (dashed curve), and $\lambda =350$ (respectively). The time evolution of $\theta _p(\tau)$ for the free Gaussian wave packet is represented by the solid curve.
}
\label{fig7}
\end{figure*}

Fig.~\ref{fig7} shows the evolution of $\theta_p(\tau)$, obtained from the Adomian series solution, for a Gaussian wave packet with $\tilde{a}=1$ (describing the effect of the background gravitational field), and for different values of the dimensionless dark energy parameter, $\lambda$. The presence of the gravitational field and of the dark energy significantly modifies the behavior of $\theta _p$. Although, in the absence of self-gravity, the peak probability density of the Gaussian wave packet satisfies $\lim _{\tau \rightarrow \infty}\theta _p^{(0)}=\infty$ in the presence of an extremely high dark energy density, corresponding to very large values of $\lambda$, the presence of self-gravitational interaction significantly alters the behaviour of $\theta _p$, at least at the first order of approximation, which may now tend to zero for finite values of $\tau$. This represents the regime in which the total collapse of the wave function occurs under the action of self-gravitational attraction, which successfully counteracts both dark energy repulsion and canonical quantum diffusion. In the range $-1\leq \tau \leq 1$, the time evolution of $\theta_p^{(1)}$ closely follows, on a qualitative level, the dynamics of $\theta_p^{(0)}$, even though some quantitative differences do appear.

\begin{figure*}[tbp!]
\centering
\includegraphics[width=7cm]{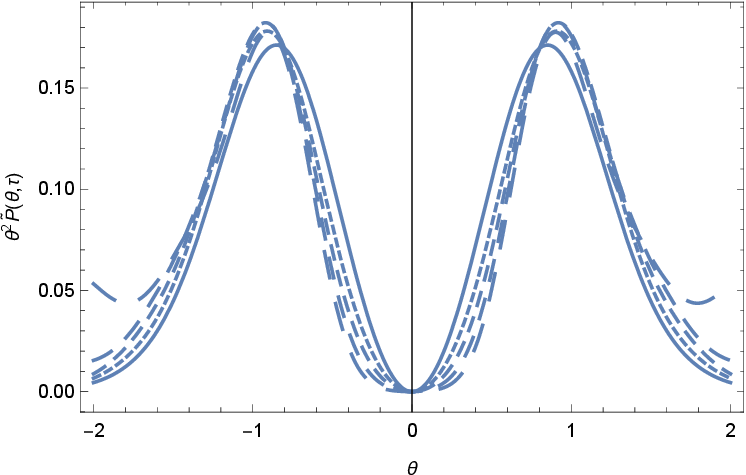}
\includegraphics[width=7cm]{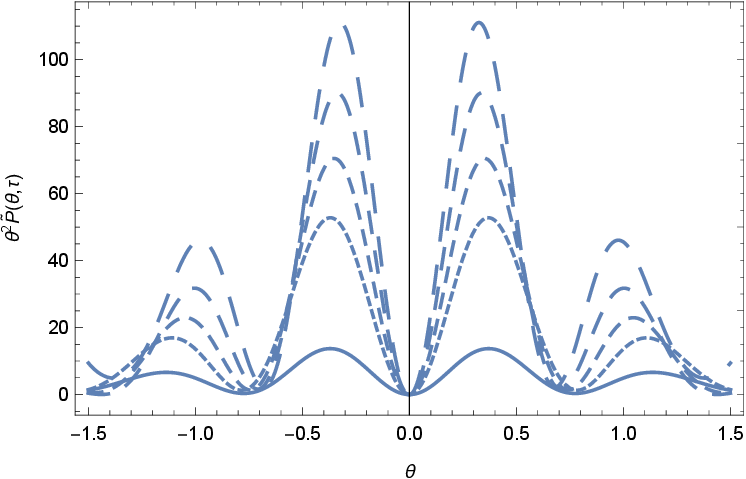}
\caption{Variation, with respect to $\theta$, of the radial probability density $\theta ^2\tilde{P}(\theta, \tau)$, for a fixed $\tau =0.25$ (left-hand panel), and for $\tau =1$ (right-hand panel), for $\tilde{a}=0$, $\sigma =0$ and $\lambda =0$, corresponding to the free Gaussian wave packet (solid curve), and for $\tilde{a}=1$, $\sigma=1$, and for different values of  $\lambda$: $\lambda =5$ (dotted curve), $\tilde{a}=1$, $\lambda =7$ (short dashed curve), $\tilde{a}=1$, $\lambda =8$ (dashed curve), and $\tilde{a}=1$, $\lambda =9$ (long dashed curve), respectively.}
\label{fig9}
\end{figure*}

In Figs.~\ref{fig9}, the dimensionless radial probability density $dP/d\theta = 4\pi \theta^2 |\tilde{\psi}|^2$ is plotted for fixed $\theta$, and for various values of $\tau$.
This clearly shows the formation of a collapsing inner core and an outer shell undergoing accelerated expansion.
The critical value of $\theta$ that demarcates between the two regions corresponds, to within an order of magnitude, to the classical turn-around radius of the particle mass, and is therefore consistent with the numerical results summarised in Table 1.

Finally, before concluding this section, we note that, since the Compton wavelength of the proton is of order $10^{-15}$ m, lab-based experiments for which the dark energy dominated regime (\ref{m_critical})-(\ref{r_p_critical}) is accessible require macromolecules with approximately 108 amu. This is two orders of magnitude below the estimated mass required for tests of the standard Schr{\"o}dinger-Newton equation using opto-mechanical traps \cite{Grossart}, which corresponds to the generic estimate for the onset of the semi-classical gravity regime with $\Lambda = 0$ \cite{Carl}. In other words, in terms of the mass parameter, current experiments are sufficiently precise to allow the effects of $\Lambda$ on the quantum dynamics of a macromolecule to be observed and measured. The associated length scale is $\sigma_0 \simeq 1-10$ m, though, unfortunately, the associated time-scales may astronomical \cite{Matt}. However, for macromolecules with $\sim 10^{10}$ amu, the canonical quantum contribution to the peak acceleration is of the same order as the dark energy contribution for $\sigma_0 \simeq 1$ m. This raises the intriguing possibility that dark energy effects may be observable in near-future experiments on local quantum systems, though, to date, the preceding order-of-magnitude estimates seem to have been overlooked in the quantum gravity literature. Crucially, the present work shows that we may go beyond such crude estimates, to obtain detailed analytical predictions of the S-N-$\Lambda$ model under realistic experimental conditions. As a proof-of-concept, our work also shows that we may fruitfully apply the ADM to any number of competing semi-classical gravity models \cite{Karolyhazy,Diosi,Penrose}. This may be useful for a range of experimental tests, including tests of gravitationally-induced wave function collapse \cite{Minar,Bose,vanMeter}.

\section{Discussions and final remarks}\label{sect6}

In the present paper, we have investigated the semi-analytical series solutions of the time-dependent Schr\"{o}dinger-Newton-$\Lambda$ (S-N-$\Lambda$) system, which describes quantum matter in the presence of a nonlinear self-gravitational interaction and a background dark energy density. For the latter, we adopted for the simple form of a positive cosmological constant, which enters into the mathematical formalism through the modified Poisson equation. In order to solve the coupled system of S-N-$\Lambda$ equations, we used a powerful mathematical method called the Adomian Decomposition Method (ADM), which provides in a fast and efficient way of obtaining series solutions of strongly nonlinear differential equations. The starting point of this method is the transformation of the given system of differential equations into an equivalent system of integral equations. Then, by positing the existence of series solutions of the integral system, one can obtain sets of recurrence relations for each unknown term in the power series expansion. 

Usually, the ADM series converges fast, allowing detailed studies of the solutions of highly nonlinear differential equations using purely analytical methods. The main advantage of the method outlined in this paper is that it is based on a rigorous mathematical procedure, namely, the series expansions of the wave function, and of the nonlinear self-gravity term, while at the same time providing results that are mathematically simple and physically intuitive. This allows the in-depth investigation of the role dark energy may play in the microscopic dynamics of a quantum particle.

In the cosmological context, the dark energy density can be inferred from the critical density of the Universe, given by $\rho _{\rm cr}=3H_0^2/8\pi G=1.88h^2\times 10^{-29}$ g/cm$^3$, where $H_0$ is the present day value of the Hubble constant, and $h=H_0/100$ km s$^{-1}$ Mpc$^{-1}$. Since the cosmological data indicates a dark energy density of the order of $\rho _{\rm vac}\simeq 0.75 \rho_{\rm cr}$, it follows that $\rho _{\rm vac}\simeq 10^{-29}$ g/cm$^3$. On the other hand, the cosmological dark energy can be obtained from physical considerations, once it is interpreted as a vacuum energy, as $\rho _{\rm vac}=\int_{k_{\rm dS}}^{\sqrt{k_{\rm Pl}k_{\rm dS}}}{\sqrt{k^2+(mc/\hbar)^2}dk}$, where $k_{\rm Pl} = 2\pi/l_{\rm Pl}$, and $k_{\rm dS} = 2\pi /l_{\rm dS}$, where $l_{\rm Pl} =\sqrt{\hbar G/c^3}$ is the Planck length and $l_{\rm dS} =\sqrt{3/\Lambda}$ is the de Sitter length \cite{Burikham:2015nma,Lake:2017ync,Lake:2017uzd}. This is consistent with the existence of the GUP and EUP \cite{Lake2020-1} and with the recent tentative observational evidence for the granular nature of dark energy on scales of order $(k_{\rm Pl}k_{\rm dS})^{-1/2} \simeq 0.1$ mm \cite{Hashiba:2018hth,Shubham2020,Perivolaropoulos:2016ucs,Antoniou:2017mhs,0.1mm_latest}. 
 
In quantum physics, a quantum fluctuation (also called vacuum fluctuation), is the random variation of the energy at a point in space, due to the creation of virtual particle-antiparticle pairs. These pairs are continuously created in the space, according to the energy-time uncertainty principle, $\Delta E \Delta t\geq \hbar/2$. In our present approach, we describe the effects of these processes on the quantum dynamics of the particle via a constant term. Even though, on a cosmological scale, the vacuum energy may have a very low (but extremely important) numerical value, quantum fluctuations may still have a significant impact on the local particle dynamics, at a microscopic level, over sufficiently long time-scales \cite{Matt}. However, as a future extension of our current work, it would be interesting to reanalyze the problem using an alternative dark energy ansatz, which captures the oscillating, or `granular' nature of the dark energy density proposed in recent models \cite{Burikham:2015nma,Lake:2017ync,Lake:2017uzd,Lake2020-1,Hashiba:2018hth,Shubham2020,Perivolaropoulos:2016ucs,Antoniou:2017mhs,0.1mm_latest}.

With or without a dark energy term, the importance of the self-gravitational interaction essentially depends on the mass of the particle. For a particle with a mass of the order of $m=10^{10}m_p\simeq 10^{-14}$ g, where $m_p$ is the proton mass, the dimensionless coefficient $\sigma$ given by Eq.~(\ref{63}) is of order unity, $\sigma \simeq 1$. In this regime, the self-gravitational interaction has a significant effect on the evolution of the quantum wave packet.
In the absence of dark energy, $\lambda = 0$, it follows from Eq.~(\ref{63l}) that this phase corresponds to the standard gravity-dominated regime of the Schr{\" o}dinger-Newton system.

In summary, the consistency of the Adomian series solutions with the exact numerical solutions obtained in previous studies represents a huge step forward in the study of the S-N-$\Lambda$ system.
 Up to now, only very crude and approximate analytical methods could be used to investigate its dynamics.
 Although useful for developing our physical intuition and providing order-of-magnitude estimates, these are no substitute for accurate quantitative solutions.
 Conversely, obtaining accurate numerical solutions is resource intensive, requiring long periods of time to develop and run the relevant codes, which are also computationally demanding \cite{Matt}.

 By contrast, the same results can be obtained using Adomian decomposition in a fraction of the time, with the help of a relatively simple Mathematica or Maple worksheet.
 Indeed, in \cite{Matt}, it was stated that ``we must deal with a complicated integro-differential equation, with little hope for analytical exploration''.
 We have now shown that this is not the case and that the S-N-$\Lambda$ system may be investigated analytically, to any degree of desired accuracy, using the right series solution techniques.
 By applying the Adomian decomposition method to PDEs, it should even be possible to obtain non-spherically symmetric solutions of the S-N-$\Lambda$ equations.
 To the best of the author's knowledge, this has not yet been attempted in the existing literature, even numerically.

 The preliminary results presented here indicate that the Adomian Decomposition Method can be used to obtain accurate solutions of a wide variety of semi-classical gravity models, subject to a wide range of initial conditions.
 Ultimately, this should help us to test the predictions of these models in greater detail, under realistic experimental conditions \cite{Grossart}.

\section*{Acknowledgments}

We thank the three anonymous reviewers for comments and suggestions that helped us to improve our manuscript. The work of TH is supported by a grant of the Romanian Ministry of Education and Research, CNCS-UEFISCDI, project number PN-III-P4-ID-PCE-2020-2255 (PNCDI III). ML thanks the Frankfurt Institute for Advanced Studies, for hospitality during the preparation of the first draft of this manuscript, the Department of Physics and Materials Science, Faculty of Science, Chiang Mai University, for providing research facilities and the Natural Science Foundation of Guangdong Province, which supported this work through grant no. 008120251030. We thank L{\' a}szl{\' o} Jenkovski, for inviting us to submit to the special is BGL-2022, and Symmetry MDPI, for 100\% waiver of the APC.


\end{document}